\documentclass[%
 reprint,
 superscriptaddress,
 amsmath,amssymb,
 aps,
 pra,
]{revtex4-1}

\usepackage{graphicx}
\usepackage{dcolumn}
\usepackage{bm}
\usepackage{hyperref}
\usepackage{url}
\usepackage{comment}
\usepackage{xspace}
\usepackage{mathtools}
\usepackage[title]{appendix}

\newcommand{\eref}[1]{Eq.~(\ref{#1})}
\newcommand{\Eref}[1]{Equation~(\ref{#1})}
\newcommand{\fref}[1]{Fig.~\ref{#1}}
\newcommand{\Fref}[1]{Figure~\ref{#1}}

\newcommand{\appropto}{\mathrel{\vcenter{
  \offinterlineskip\halign{\hfil$##$\cr
    \propto\cr\noalign{\kern2pt}\sim\cr\noalign{\kern-2pt}}}}}

\newcommand\Ne{$n_\mathrm{e}$\xspace}
\newcommand\Te{$T_\mathrm{e}$\xspace}
\newcommand\Xlev{$\mathrm{X^1\Sigma_g^+}$\xspace}

\newcommand\alev{$\mathrm{a^3\Sigma_g^+}$\xspace}
\newcommand\clevp{$\mathrm{c^3\Pi_u^+}$\xspace}

\newcommand\clevpm{$\mathrm{c^3\Pi_u^\pm}$\xspace}
\newcommand\blev{$\mathrm{b^3\Sigma_u^+}$\xspace}

\newcommand\dlevm{$\mathrm{d^3\Pi_u^-}$\xspace}
\newcommand\dlevpm{$\mathrm{d^3\Pi_u^\pm}$\xspace}

\newcommand\ilevpm{$\mathrm{i^3\Pi_g^\pm}$\xspace}

\newcommand\jlevpm{$\mathrm{j^3\Delta_g^\pm}$\xspace}
\newcommand\elev{$\mathrm{e^3\Sigma_u^+}$\xspace}
\newcommand\hlev{$\mathrm{h^3\Sigma_g^+}$\xspace}
\newcommand\glev{$\mathrm{g^3\Sigma_g^+}$\xspace}

\newcommand\Blev{$\mathrm{B^1\Sigma_u^+}$\xspace}

\newcommand\Clevpm{$\mathrm{C^1\Pi_u^\pm}$\xspace}

\newcommand\EFlev{$\mathrm{EF^1\Sigma_g^+}$\xspace}
\newcommand\Dlevp{$\mathrm{D^1\Pi_u^+}$\xspace}
\newcommand\Dlevm{$\mathrm{D^1\Pi_u^-}$\xspace}
\newcommand\Dlevpm{$\mathrm{D^1\Pi_u^\pm}$\xspace}
\newcommand\Bplev{$\mathrm{B'^1\Sigma_u^+}$\xspace}
\newcommand\GKlev{$\mathrm{GK^1\Sigma_g^+}$\xspace}

\newcommand\Ilevpm{$\mathrm{I^1\Pi_g^\pm}$\xspace}

\newcommand\Jlevpm{$\mathrm{J^1\Delta_g^\pm}$\xspace}
\newcommand\Hlev{$\mathrm{H\bar{H}^1\Sigma_g^+}$\xspace}

\newcommand{\mycomment}[1]{}

\begin{document}
\newcommand{\ManuscriptTitle}{
    Experimental Validation of Collision-Radiation Dataset 
    for Molecular Hydrogen in Plasmas
}

\title{\ManuscriptTitle}


\author{Keisuke Fujii}
\email{fujiik@ornl.gov}
\affiliation{%
    Fusion Energy Division, Oak Ridge National Laboratory, Oak Ridge, TN 37831-6305, United States of America
}
\author{Keiji Sawada}
\affiliation{%
    Faculty of Engineering, Shinshu University, 4-17-1 Wakasato, Nagano 380-8553, Japan
}
\author{Kuzmin Arseniy}
\affiliation{%
    Department of Mechanical Engineering and Science,
    Graduate School of Engineering, Kyoto University
    Kyoto 615-8540, Japan
}
\author{Motoshi Goto}
\author{Masahiro Kobayashi}
\affiliation{%
    National Institute for Fusion Science, Toki, Gifu, 5909-5292, Japan
}
\author{Liam H. Scarlett}
\author{Dmitry V. Fursa}
\author{Igor Bray}
\affiliation{%
    Department of Physics and Astronomy, 
    Curtin University, Perth, Western Australia 6102, Australia
}
\author{Mark C. Zammit}
\affiliation{%
    Theoretical Division, Los Alamos National Laboratory, Los Alamos, NM 87545, United States of America
}
\author{Theodore M. Biewer}
\affiliation{%
    Fusion Energy Division, Oak Ridge National Laboratory, Oak Ridge, TN 37831-6305, United States of America
}

\date{\today}

\begin{abstract}
    Quantitative spectroscopy of molecular hydrogen has generated substantial demand, leading to the accumulation of diverse elementary-process data encompassing radiative transitions, electron-impact transitions, predissociations, and quenching. 
    However, their rates currently available are still sparse and there are inconsistencies among those proposed by different authors.
    In this study, we demonstrate an experimental validation of such molecular dataset by composing a collisional-radiative model (CRM) for molecular hydrogen and comparing experimentally-obtained vibronic populations across multiple levels.
    
    From the population kinetics of molecular hydrogen, the importance of each elementary process in various parameter space is studied.
    In low-density plasmas (electron density $n_\mathrm{e} \lesssim 10^{17}\;\mathrm{m^{-3}}$) the excitation rates from the ground states and radiative decay rates, both of which have been reported previously, determines the excited state population.
    The inconsistency in the excitation rates affects the population distribution the most significantly in this parameter space.
    On the other hand, in higher density plasmas ($n_\mathrm{e} \gtrsim 10^{18}\;\mathrm{m^{-3}}$), the excitation rates \textit{from} excited states become important, which have never been reported in the literature, and may need to be approximated in some way.
    
    In order to validate these molecular datasets and approximated rates, we carried out experimental observations for two different hydrogen plasmas; 
    a low-density radio-frequency (RF) heated plasma ($n_\mathrm{e}\approx 10^{16}\;\mathrm{m^{-3}}$) and the Large Helical Device (LHD) divertor plasma ($n_\mathrm{e}\gtrsim 10^{18}\;\mathrm{m^{-3}}$).
    The visible emission lines from 
    \EFlev, \Hlev, \Dlevpm, \GKlev, \Ilevpm, \Jlevpm, 
    \hlev, \elev, \dlevpm, \glev, \ilevpm, and \jlevpm states were observed simultaneously and their population distributions were obtained from their intensities.
    We compared the observed population distributions with the CRM prediction, in particular the CRM with the rates compiled by 
    Janev et al.,
    Miles et al., and
    and those calculated with the molecular convergent close-coupling (MCCC) method.
    The MCCC prediction gives the best agreement with the experiment, particularly for the emission from the low-density plasma.
    On the other hand, the population distribution in the LHD divertor shows a worse agreement with the CRM than those from low-density plasma, indicating the necessity of the precise excitation rates from excited states.
    We also found that the rates for the electron-attachment is inconsistent with experimental results. This requires further investigation.
\end{abstract}


\clearpage 

\maketitle

\section{Introduction}

Molecular hydrogen emission appears in various low-temperature plasmas, such as interstellar media~\cite{Rosenthal2000-yb}, process plasmas~\cite{Lieberman2005-ac}, 
and cold-temperature regions of fusion plasmas~\cite{Janev2003-ca,Fantz2001-qf,ishihara}.
In particular, in fusion devices, hydrogen (or its isotopologues) molecules play an important role in the particle balance chain~\cite{Stangeby2000-fx}, as well as one of efficient heat exhaust by molecular assisted processes, such as molecular-assisted recombination~\cite{Janev2003-ca,Ohno1998-ir,Verhaegh2022-kn}.
Despite its importance, its quantitative diagnostics have been rather unestablished. 

A key technique to bridge the experimentally observable emission lines and particle (ions, atoms, and molecules) behavior in plasmas is \textit{collisional-radiative model} (CRM), which solves the excited-state population based on the rate equations.
Since CRMs require a complete dataset of elementary process rates, such as the electron-impact transition and spontaneous decay, a huge amount of effort has been paid to measure, calculate, compile, and validate these rates.
Thanks to the efforts, the recommended rates for atomic species have been compiled in a database, e.g., ADAS~\cite{ADAS}, and have been in active use for diagnostic purposes.

However, the situation for molecules is behind that for atomic species.
The essential difficulty of molecular spectroscopy is in the complex quantum structure of molecules; the interplay between electron and nuclei motion makes the energy structure more complex.
This complexity leads to a higher hurdle in composing the elementary-process rates, both experimentally and theoretically.
Hydrogen molecule, which is the simplest neutral molecule, has been studied for a long time, and many rates have been reported, but there is still inconsistency between the results among authors (see Sec.~\ref{subsec:excitation} later). 
CRMs for hydrogen molecules have also been developed by several authors based on these rates~\cite{Sawada1995-gg,Greenland2002-cr,Lavrov2006-bu,Guzman2013-vt,Shakhatov2016-ci,Sawada2016-zf,Wunderlich2021-td,yacora}, however, their validity remains unclear and an experimental evidence to solve the inconsistencies are still lacking.

The purpose of this paper is to compose a CRM for molecular hydrogen based on the currently-available datesets and to validate these dataset by comparing our CRM predictions with experimental observations. 

In the next section, we will present the basic principle of the CRM and the important elementary processes for hydrogen molecule.
Inconsistencies among the datasets reported by different authors, such as those proposed by Miles~\cite{Miles1972-md}, Janev~\cite{Janev2003-ca}, and those theoretically calculated with the recently-developed molecular convergent close-coupling (MCCC) method~\cite{mccc,Scarlett2017,Scarlett2021-ag,Scarlett2021-gc,Scarlett2021-oi,Scarlett2021-pc,Scarlett2021-pc,Scarlett2022-zj,Scarlett2022-zj,Scarlett2023-co} will be pointed out.
Based on our CRM, in Sec.~\ref{sec:crm_dynamics}, we will discuss the population kinetics and its dependence on the plasma parameters. 
In particular, we will show that the population kinetics is essentially different in low-density ($n_\mathrm{e} \lesssim 10^{17}\;\mathrm{m^{-3}}$, \textit{coronal} phase) and high-density regions ($n_\mathrm{e} \gtrsim 10^{18}\;\mathrm{m^{-3}}$, \textit{saturation} phase).
The important elementary processes are also different in these phases.
Finally, in Sec.~\ref{sec:experiment} we demonstrate the experimental validation of our CRM and point out the current limitation in molecular-hydrogen spectroscopy.

\section{CRM for Molecular Hydrogen and the Elementary Processes\label{sec:crm}}

\begin{figure}[tbp]
    \includegraphics[width=8cm]{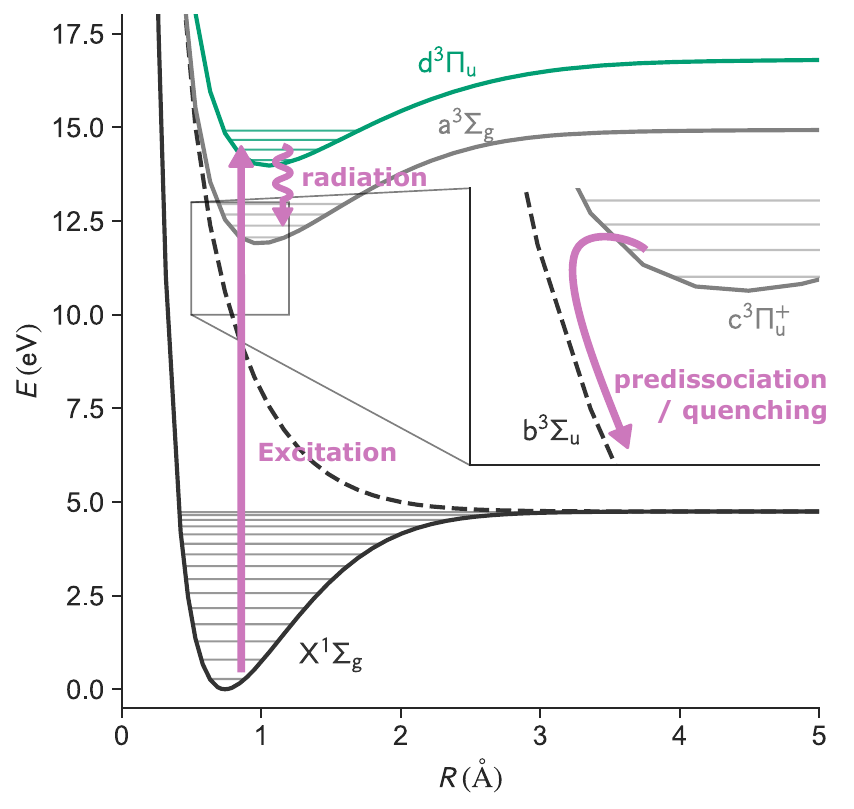}
    \caption{%
        A schematic illustration of the elementary processes of molecules in plasmas.
    }
    \label{fig:processes}
\end{figure}

A CRM is essentially a set of rate equations for the excited-state population with various excitation / deexcitation processes taken into account.
For molecular hydrogen, the important processes are 
radiative decay (Sec.~\ref{subsec:radiation}), 
electron-impact excitation, deexcitation, ionization (Sec.~\ref{subsec:excitation}), 
predissociation (Sec.~\ref{subsec:predissociation}), quenching (Sec.~\ref{subsec:quenching}), and electron attachment (Sec.~\ref{subsec:attachment}).
These processes are schematically illustrated in \fref{fig:processes}.

By including these processes, the rate equation for the population in the excited state $p$ can be written as follows:
\begin{align}
    \label{eq:crm_population}
    \frac{d}{dt}n_p 
    &= \sum_{q} [C_{p \leftarrow q} n_e + A_{p\leftarrow q}] n_q \\
    \label{eq:crm_depopulation}
    &- \sum_{q} [C_{q \leftarrow p} n_e + A_{q\leftarrow p}] n_p \\
    \notag
    &- \sum_{k} [C_{k \leftarrow p} n_e + C_{k \leftarrow p}^\mathrm{attach} n_e  + \\
    \label{eq:crm_loss}
    & \;\;\;\;\;\;\;\;\;\;\;\;
    A_{k\leftarrow p} + P_{k\leftarrow p} + Q_{k\leftarrow p} n_\mathrm{H_2}] n_p,
\end{align}
where the first term on the right hand side (\eref{eq:crm_population}) is the population influx from $q$ state to $p$ state, while the second (\eref{eq:crm_depopulation}) and third terms (\eref{eq:crm_loss}) are the population outflux from $p$ state to $q$ and $k$ state, respectively.
Here, $q$ indicates stable states of molecular hydrogen, and $k$ indicates other states, such as dissociative unstable states and bound states of molecular ions.
For the sake of simplicity, we assume the ionizing plasma, so that the recombination is not important. 
This means that there is no influx \textit{from} the dissociative and ionized states.
This assumption may be valid since the volume association and recombination are often negligible in the density range considered here (\Ne$\lesssim 10^{20}\mathrm{\;m^{-3}}$), since the dominant generation process of molecules is the surface-assisted association. 

$A_{p \leftarrow q}$ indicates the radiative decay rate, while $P_{k\leftarrow p}$ is the predissociation rate, both of which are spontaneous processes.
$C_{p \leftarrow q}$ indicates the rate coefficient by electron impact (excitation, deexcitation, and ionization). 
$C_{k \leftarrow q}^\mathrm{attach}$ represents the rate coefficient of the electron-attachment process
\begin{align}
    \label{eq:attachment}
    \mathrm{H}_2^\star + e^- \rightarrow \mathrm{H}^- + \mathrm{H}
\end{align}
in which the electron will attach to an excited molecule eventually leading to the dissociation.

$Q_{k\leftarrow p}$ is the rate coefficient for the quenching process,
\begin{align}
    \label{eq:quenching}
    \mathrm{H}_2^\star + \mathrm{H}_2 \rightarrow 2\mathrm{H} + \mathrm{H}_2,
\end{align}
which also leads a dissociation eventually.

By assuming the quasi steady state for all the excited-state population $dn_p / dt = 0$, we obtain the population density
\begin{align}
    \label{eq:crm_solution}
    n_p = R_1^p(n_e, T_e) n_e n_1,
\end{align}
where $n_1$ is the ground state density, and $R_1^p(n_e, T_e)$ is called collisional-radiative population coefficient.

\subsection{Quantum structure of hydrogen molecule\label{subsec:structure}}

\begin{figure*}[tbh]
    \includegraphics[width=17cm]{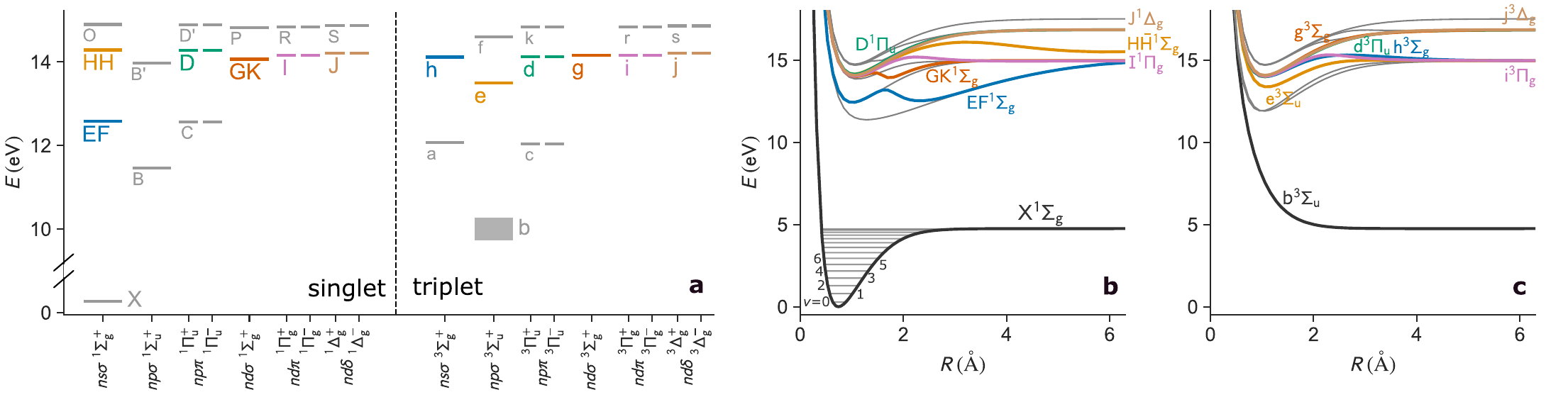}
    \caption{%
        (a) A simplified level diagram of molecular hydrogen. 
        (b), (c) Potential energy curves for the hydrogen molecules. The vibrational levels are indicated by horizontal bars for \Xlev states.
    }
    \label{fig:potential}
\end{figure*}

The energy structure of molecules is dominantly represented by three types of motions: electron motion, nuclear vibration, and nuclear rotation.
Since the time scale of the electron motion ($\approx 10^{-15}$ s) is faster than the nuclear vibration ($\approx 10^{-13}$ s) and nuclear rotation ($\approx 10^{-12}$ s), the nuclei rarely move while an electron rotates around them.
In the Born-Oppenheimer approximation, we simply solve the Sch\"oldinger equation only for electrons with fixed nuclear positions, based on the time-scale difference of their motion.

\Fref{fig:potential} (a) shows a level diagram of the electronic structure of molecular hydrogen.
On one hand, since two electrons are orbiting around the nuclei in a hydrogen molecule, the electron energy structure has some similarity with the energy structure of atomic helium; there are two energy manifolds depending on the relative directions of two electron spins, i.e., singlet ($S=0$) and triplet ($S=1$) states.
Here, $S$ indicates the spin quantum number of the electrons.
On the other hand, the axially-symmetric potential of a diatomic molecule violates the conservation of the angular momentum of the electrons, but instead its amplitude of the projection to the molecular axis is conserved.
A quantum number $\Lambda = |M_L|$ is assigned to this projection and the states with different values of $\Lambda$ have different energies, where $L$ is the orbital quantum number for electron, and $M_L$ is the projection to the molecular axis.
Capital greek letters $\Sigma, \Pi, \Delta, \ldots$ are assigned to $\Lambda=0, 1, 2, \ldots$ states, respectively (see the horizontal axis in \fref{fig:potential}~(a)).
For the states with $\Lambda > 0$, there are twofold degeneracies, corresponding to $M_L = -\Lambda$ and $M_L = +\Lambda$ states.
This degeneracy is related to the symmetry property of the electron wavefunction.
The electron wavefunction must be either symmetric or anti-symmetric against the reflection at any plane passing through both nuclei.
If the sign of the wavefunction remains unchanged by this reflection, a superscript $+$ is assigned to this state, while for the other case a superscript $-$ is assigned. 

For the diatomic molecules with the same-charge nuclei (e.g., $\mathrm{H_2}$ and HD), the electron wavefunction has another symmetry property, where the wavefunction is either symmetric or anti-symmetric against the reflection around the center of the two nuclei.
If the sign of the wavefunction remains unchanged by this reflection $g$ is assigned to this state, for while for the other case $u$ is assigned (from the German \textit{gerade} and \textit{ungerade}). 

Conventionally, a capital letter X is assigned to the electronic ground state, and B, C, $\dots$ are used for the 1st, 2nd, $\dots$ electronic excited states in the same multiplet to the X state, while lower case letters a, b, $\dots$ are used for the other multiplet states [
for historical reasons, there are some irregularities in the labelling of hydrogen molecule states in these conventional names, such as the EF state and B' state, as shown in \fref{fig:potential} (b)].
Each electronic state has been specified by the notation $\mathrm{[assigned\;letter]} \;^{(2S+1)}\Lambda_\mathrm{[g\; or \;u]}^{+/-}$.
To avoid the confusion, in this paper, we use full notation for each electronic state, e.g., \EFlev and \dlevm.

The Born-Oppenheimer approximation gives the excited energy of electrons with a fixed nuclear position. 
Its dependence on the inter-nuclear distance can be interpreted as the potential surfaces for nuclear motion.
\Fref{fig:potential}~(b) and (c) show the potential energy curves as a function of the internuclear distance $R$ for the singlet and triplet states, respectively~\cite{Nakashima2018-ip,Wolniewicz1993-av}.
\blev state in the triplet does not have a local minimum, meaning a purely dissociative state.
The vibrational and rotational motions of nuclei are obtained by solving the Schr\"odinger equation for nuclear motions with these potential energy curves. 
In \fref{fig:potential}~(b), the vibrational energy levels for \Xlev are indicated by horizontal lines.
The vibrational and rotational energy intervals for hydrogen molecules are $\approx 0.5$ eV and $\approx$ 0.05 eV, respectively.
In our CRM, only the electronic and vibrational states will be considered, and the rotational levels are ignored. 
On the other hand, the rotational states are needed to be taken into account in the analysis of an observed emission spectrum. 

\subsection{Radiative transition\label{subsec:radiation}}

\begin{figure*}[tbp]
    \includegraphics[width=15cm]{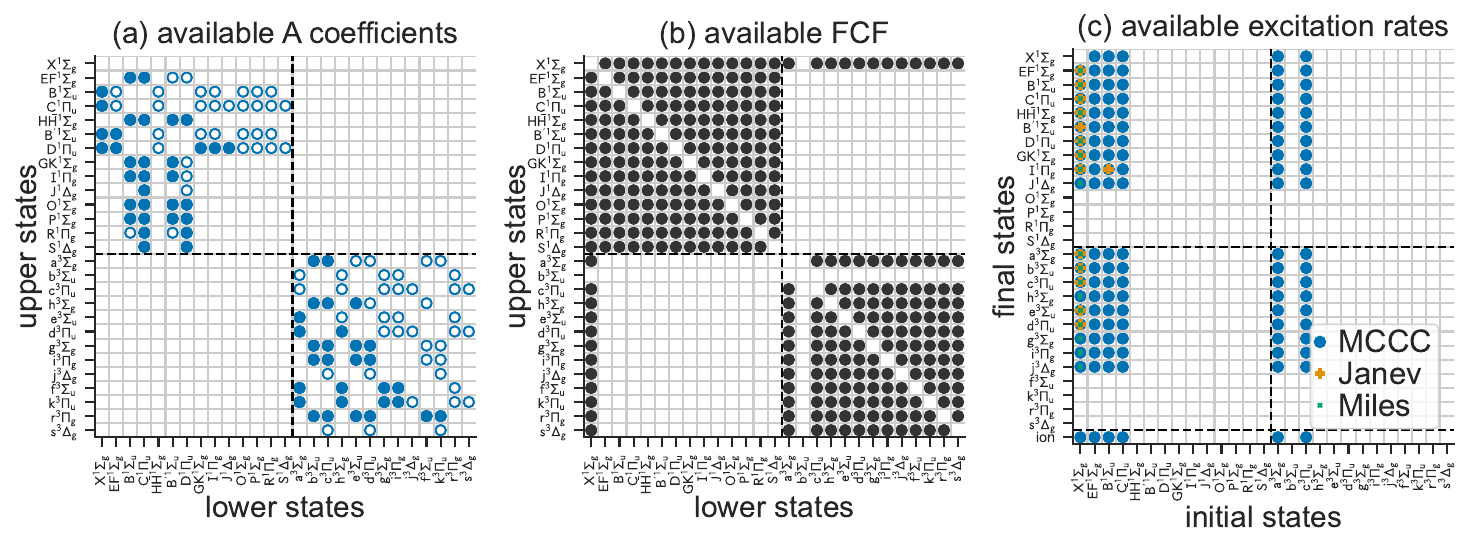}
    \caption{%
        Transitions available in each dataset. (a) Radiative decay rates ($A$) by Fantz et al.~\cite{Fantz2006-cn}. Filled markers indicate the available data, while the open markers indicate the optically-allowed transitions but the data is not available.
        (b) the Franck-Condon factors (FCF) by Fantz et al.~\cite{Fantz2006-cn}. Note that the Franck-Condon factors are symmetric, i.e., $q_{\alpha'v'-\alpha''v''} = q_{\alpha''v''-\alpha'v'}$.
        (c) Availability of the electron-impact excitation cross-sections. 
        The circle markers indicate the available data in MCCC database~\cite{mccc}, while plus markers and dots indicate those available in Janev dataset~\cite{Janev2003-ca} and Miles~\cite{Miles1972-md}, respectively.
    }
    \label{fig:selection_rule}
\end{figure*}

As similar to an atomic system, an excited molecule decays to a lower state by emitting a photon.
The selection rules for the electronic states is
$|\Lambda' - \Lambda''| \leq 1$, $S' - S'' = 0$, and $\mathrm{g \leftrightarrow u}$.
\Fref{fig:selection_rule}~(a) shows the allowed transitions among each state by markers.
Note that according to the convention, a single prime $'$ and double prime $''$ are used to indicate the initial and final states of each transition, respectively. 

The radiative decay rate among electronic and vibronic states, $A^{\alpha' v'}_{\alpha''v''}$, have been computed by Funtz et al~\cite{Fantz2006-cn} based on the electronic dipole transition moment available in literature, not only for $\mathrm{H_2}$, but also all the possible isotopologues.
The transitions with the data available are shown by filled markers in \fref{fig:selection_rule}~(a).

Among allowed transitions among the listed levels, the one for \clevpm$\leftarrow$\jlevpm is missing in their data set, because of the unavailability of the electronic dipole transition moment.
We approximate the transition rate based on the experimental value by Astashkevich et al~\cite{Astashkevich1996-ib}, and the method based on Franck-Condon factor and H\"onl-London factor.

\subsection{Electron-impact transition\label{subsec:excitation}}

\begin{figure*}
    \includegraphics[width=17cm]{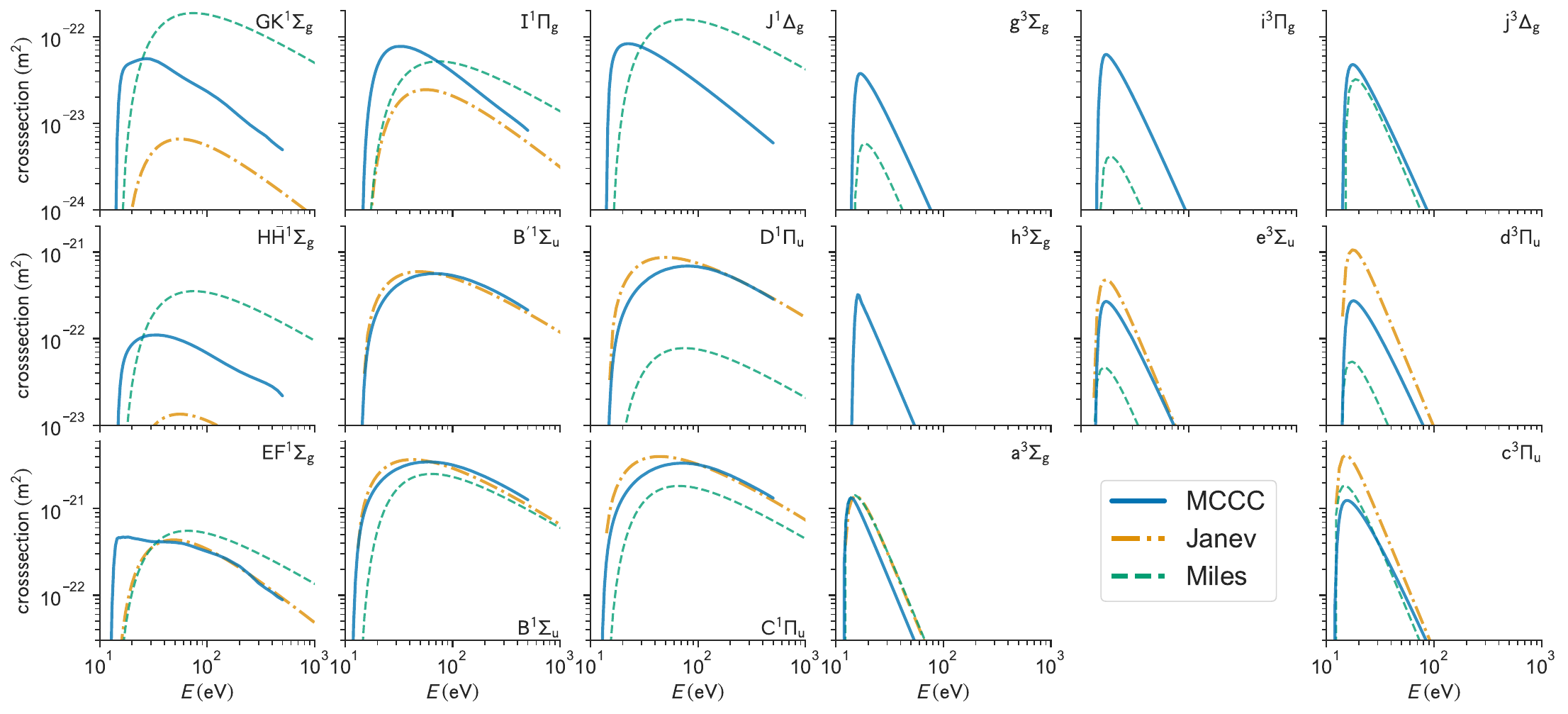}
    \caption{%
        A comparison of the excitation cross sections from \Xlev$(v=0)$ state to the $n=2$ and $n=3$ excited states. 
        The vibrational quantum number for final state are not distinguished (the cross-sections are summed up).
        The cross-sections by MCCC~\cite{mccc}, Janev~\cite{Janev2003-ca}, and Miles~\cite{Miles1972-md} are shown by different lines.
    }
    \label{fig:crosssections}
\end{figure*}

Electron-impact transition is the most important process that generates excited-state molecules in many plasmas. 
Several groups have proposed different sets of cross-sections.
Miles et al. have assumed an analytical form for the vibrationally-resolved electron-impact cross sections based on the generalized oscillator strength method, and adjust a few parameters so that this form fits experimental data available at that time~\cite{Miles1972-md}.
Later, Janev et al. have also compiled various electron-impact cross sections for molecular hydrogen~\cite{Janev2003-ca}.
However, there has been inconsistency among them.
As seen in \fref{fig:crosssections}, a significant difference can be seen in some of the cross sections, such as \Xlev $\rightarrow$ \Hlev.

Also recently, a systematic theoretical method based on MCCC method has been developed which is able to calculate the vibrational-state-resolved electron-impact transition cross-sections~\cite{mccc, Scarlett2017,Scarlett2021-ag,Scarlett2021-gc,Scarlett2021-oi,Scarlett2021-pc,Scarlett2021-pc,Scarlett2022-zj,Scarlett2022-zj,Scarlett2023-co}.
The MCCC rates are also shown in \fref{fig:crosssections}. These three still show inconsistencies, although the experimental comparison for the \dlevm state population by W\"underlich et al. suggests a better performance of the MCCC cross sections~\cite{Wunderlich2021-td}.

MCCC also provides some of the cross sections among excited states. 
The available rates are shown in \fref{fig:selection_rule}~(c).
However, many cross sections \textit{from} excited states are still missing, which are important for high-density plasmas as described below.
The following two approximation methods for such missing rates can be considered. 

\subsubsection{Helium approximation\label{subsubsec:helium}}
In the first CR model for hydrogen molecule~\cite{Sawada1995-gg}, the authors have used the cross-section for the transition among helium excited states.
For example, \EFlev has $2s$ electron orbit, while \Dlevm has $3p$ orbit.
Thus, this rate coefficient may be approximated as follows,
\begin{align}
    C_{\mathrm{D^-}(v'') \leftarrow \mathrm{EF}(v')} \approx 
    \frac{1}{3} C^\mathrm{He}_{3 ^1\mathrm{P} \leftarrow 2 ^1\mathrm{S}} \;q_{\mathrm{EF}(v')-\mathrm{D}(v'')},
\end{align} 
where $C^\mathrm{He}_{3 ^1\mathrm{P} \leftarrow 2 ^1\mathrm{S}}$ is the excitation rate of helium atom for its $3 ^1\mathrm{P} \leftarrow 2 ^1\mathrm{S}$ transition. 
The factor $1/3$ accounts for the three $3 ^1\mathrm{P}$ states in hydrogen molecule (i.e., \Bplev, \Dlevp, \Dlevm). 
$q_{\mathrm{EF}(v')-\mathrm{D}(v'')}$ is the Franck-Condon factor between \EFlev and \Dlevpm states.

\begin{figure*}
    \includegraphics[width=17cm]{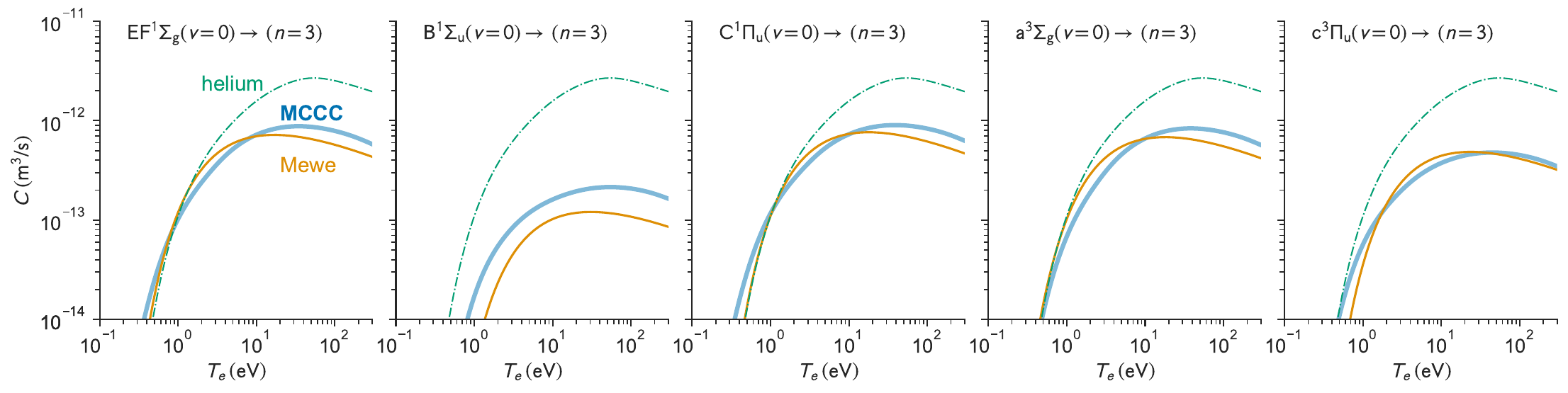}
    \caption{%
        The total rate coefficients from $n=2$ states (\EFlev, \Blev, \Clevpm, \alev, \clevpm) to all of the $n=3$ states.
        The bold curves are from the MCCC database~\cite{mccc}, while the helium approximation and Mewe's approximation are shown by chain curves and thin curves, respectively. See the text for details.
    }
    \label{fig:rate_comparison}
\end{figure*}

In order to see the accuracy of this approximation,  
we show in \fref{fig:rate_comparison} the rate coefficients from each $n=2$ states (\EFlev, \Blev, \Clevpm, \alev, \clevpm) to any of $n=3$ states.
The bold blue curves show the calculation result by MCCC, while the orange chain curves show those for helium atoms. 
These two sets of cross sections agree within a factor of $\approx 10$. 

\subsubsection{Mewe's approximation\label{subsubsec:mewe}}

Another possible approximation would be the one proposed by Mewe~\cite{Mewe_undated-qm}, based on the line-strength for dipole transitions.
With this approximation, the rate coefficient for the electron-impact excitation from $q$ state to $p$ state can be written as
\begin{align}
    \label{eq:mewe}
    C_{p\leftarrow q}(T_e)\approx 
    \frac{1}{g_q}\frac{\beta}{\sqrt{kT_e}}S_{qp}\exp\left[-\frac{\omega_{pq}}{kT_e}\right],
\end{align}
with 
\begin{align}
    \beta = \frac{2^{7/2}}{3\sqrt{3}}\pi^{3/2}
    \alpha a_0^2 c\sqrt{E_\mathrm{H}}\frac{1}{e^2a_0^2}\xi
\end{align}
where $g_p$ is the statistical weight of the initial state, $k$ is the Boltzmann constant, $T_e$ is the electron temperature, and $\omega_{pq} = E_q - E_p$ is the energy difference between states $p$ and $q$, $E_\mathrm{H} \approx$ 27.2 eV is the Hartree energy, $\alpha$ is the fine structure constant, $e$ is the elementary charge, $a_0$ is the Bohr radius, and $c$ is the light speed.
$S_{qp}$ is the line strength of the transition, which is evaluated from the A coefficients for the transition, 
\begin{equation}
    \label{eq:radiative_rate}
    A_{p \gets q} = \gamma \frac{1}{g_q}\omega_{pq}^3 S_{pq}.
\end{equation}
with
\begin{equation}
    \label{eq:gamma}
    \gamma = \frac{4}{3} \alpha^4 \frac{c}{a_0} \frac{1}{E_\mathrm{H}^3}\frac{1}{e^2 a_0^2}
\end{equation}

$\xi$ is the gaunt factor, which is represented by the following equation
\begin{align}
\xi = a + b \exp\left(\frac{\omega_{pq}}{kT_e}\right)\exp_1\left(\frac{\omega_{pq}}{kT_e}\right),
\end{align}
where $a\approx 0$ and $b\approx 1$ are parameters that depend on the particular system.
We assume $a = -0.03$ and $b = 0.28$ so that \eref{eq:mewe} best represents the MCCC rate coefficients.
Orange curves in \fref{fig:rate_comparison} show the rate coefficients by \eref{eq:mewe}. 
\Eref{eq:mewe} (orange thin curve) represents the MCCC rate coefficients better than the corresponding helium rates (green chain curve). 
This is partly because of the degrees of freedom in $\xi$ (we basically \textit{fit} by adjusting $a$ and $b$) but also because it takes into account the energy difference of the actual system.

\subsection{Predissociation\label{subsec:predissociation}}

Predissociation is a spontaneous process where an excited molecule undergoes an internal conversion to another state, typically leading to a dissociation.
One of the important dissociation paths is 
\begin{align}
    \mathrm{H}_2(c^3\Pi^+) \rightarrow \mathrm{H}_2(b^3\Sigma^+) \rightarrow 2\mathrm{H},
\end{align}
the decay rate of which is $1.6\times 10^{8}\;\mathrm{s}^{-1}$ and $7.0\times 10^{8}\;\mathrm{s}^{-1}$ for $v=0, N=1$ and $v=1, N=1$ states, respectively~\cite{Comtet1985-ep}. 
Although \clevp is a radiatively metastable state, this predissociation process reduces its lifetime significantly.

The lifetime of an isolated excited molecules $\tau$ can be written as
\begin{align}
    \frac{1}{\tau_p} = \sum_{q} A_{q\leftarrow p} + \sum_{q} P_{q\leftarrow p}.
\end{align}
The values of $\tau_p$ have been compiled by Astashkevich et al~\cite{Astashkevich2015-gd}.
We evaluate $\sum_{q} P_{q\leftarrow p}$ for each excited states by the table in their paper.
Note that although the predissociation rate weakly depends on the rotational quantum states, we use the value for the smallest $N$ state for our CR model.

\subsection{Quenching rates\label{subsec:quenching}}
Quenching is a similar process to the predissociation, but induced by a heavy-particle collision (\eref{eq:quenching}).
Wedding et al has reported the rate coefficient 
$(1.88 \pm 0.10) \times 10^{-15}\mathrm{\;m^3s^{-1}}$ for 300-K $\mathrm{H_2}$ collisions 
~\cite{Wedding1988-ul}.
As this number is $\approx 10^{2}$ times smaller than the electron-impact excitation rate (\fref{fig:rate_comparison}), this process becomes important for low-ionization-degree plasmas, $n_\mathrm{e} / n_\mathrm{H_2} \lesssim 10^{-3}$.
In our experimental conditions described below, this process is not significant.

\subsection{Dissociative attachment\label{subsec:attachment}}
In Ref.~\cite{Datskos1997-tf}, it has been suggested that the dissociative attachment (\eref{eq:attachment}) is very efficient for \textit{high-Rydberg} state of molecualr hydrogen, with the rate coefficient of $C^\mathrm{attach} \approx 6\times 10^{-11} \mathrm{\;m^{3}/s}$.
However, the original paper~\cite{Datskos1997-tf} has not specified which states are regarded as a \textit{high-Rydberg} state. 
In Refs.~\cite{Wunderlich2020-ds, yacora}, the authors have assumed this value for $n\geq 3$ states.
In our paper, we also examine the validity of this assumption.

\section{CRM predictions and population dynamics\label{sec:crm_dynamics}}

Our CRM described in this paper considers the population of molecule in each electronic- and vibrational-states up to $n=4$ states, while the rotational states are not resolved.
Also, we assume that the population in the electronic-ground state can be represented by the Boltzmann distribution,
\begin{align}
    n_\mathrm{X}(v) = n_\mathrm{H_2} \frac{1}{Z} \exp\left(-\frac{E_v}{k T_v}\right),
\end{align} 
where $\mathrm{H_2}$ is the molecular density in \Xlev state, $E_v$ is the energy of vibrational state $v$ in \Xlev, and $T_v$ is the vibrational temperature. $k$ is the Boltzmann constant.
Here, $Z$ is the normalization constant,  
\begin{align}
    Z = \sum_v \exp\left(-\frac{E_v}{k T_v}\right).
\end{align} 
The solution of the CRM can be represented by a similar manner to \eref{eq:crm_solution}, but
\begin{align}
    \label{eq:crm_solution_H2}
    n_p = R_1^p(n_e, T_e, T_v) n_e n_\mathrm{H_2}.
\end{align} 

\begin{figure*}
    \includegraphics[width=18cm]{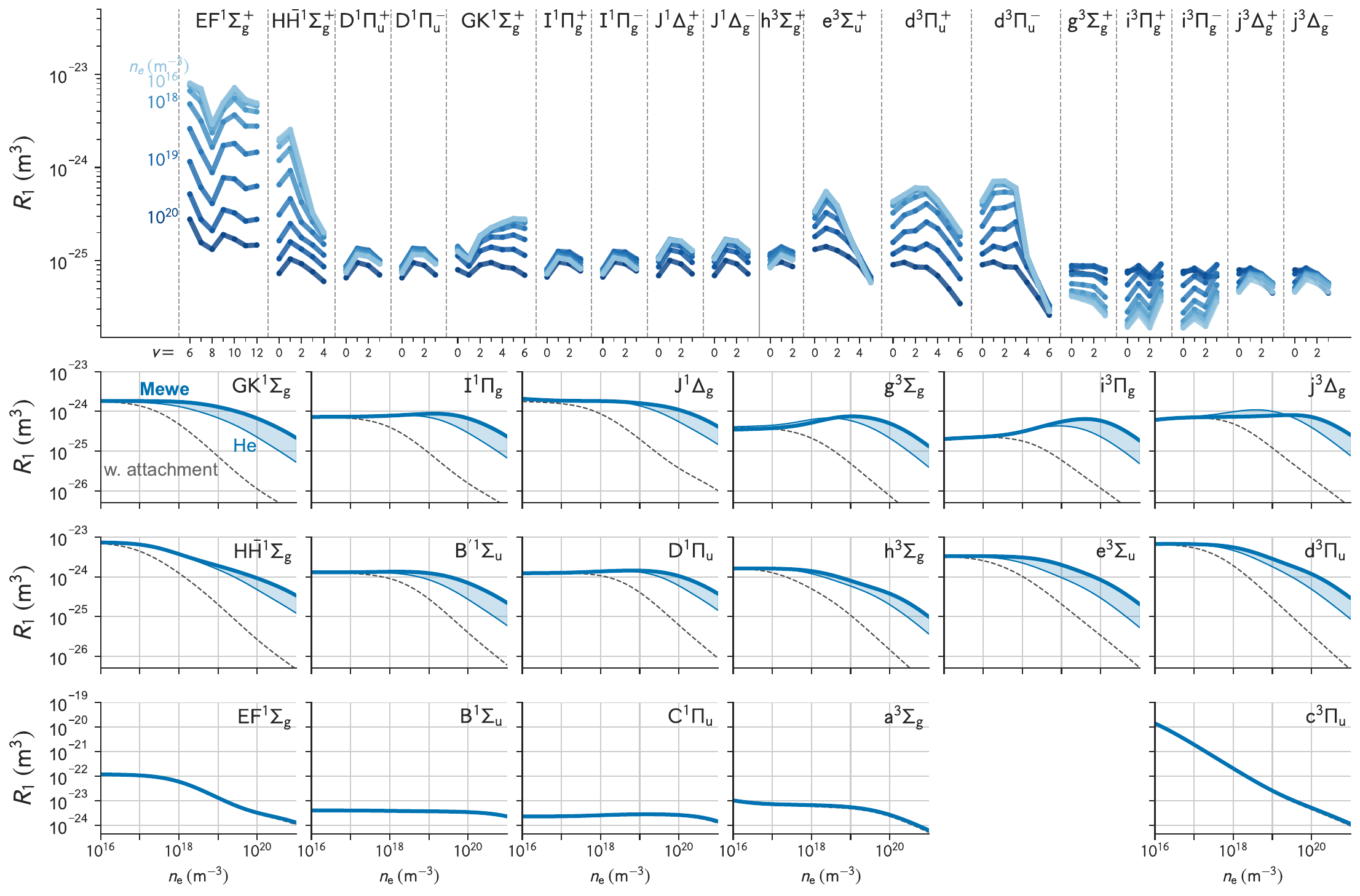}
    \caption{%
        (Top panel) Population coefficients $R_1^p$ for some excited states of molecular hydrogen calculated with our CRM, and its \Ne-dependence.
        $T_\mathrm{e} = 10$ eV and $T_v = 0.1$ eV are assumed.
        $R_1$ for \EFlev has a significant \Ne-dependence, while that for \Dlevpm is small.
        (2nd, 3rd, and 4th rows): \Ne-dependence for $n=2$ and $n=3$ states. The population for the vibrational states are summed up.
        The thick solid curves show the population coefficients calculated with MCCC cross-sections and Mewe's approximations for the missing transitions, 
        while thin solid curves show the population coefficients calculated with helium cross-sections for missing transitions.
        These two shows a significant difference in the high-\Ne space, which is the systematic uncertainty in the current CRM.
        Thin dotted curve shows the result with the dissociative attachment included.
        Since the rate coefficient for the dissociative attachment recommended in Ref.~\cite{yacora} is large compared with the electron-impact rate, the inclusion of this process significantly changes the population. 
        In particular, the critical density between the \textit{coronal} phase (where $R_1^p \approx n_\mathrm{e}^0$) and the \textit{saturation} phase (where $R_1^p \approx n_\mathrm{e}^{-1}$) moves to the lower density side.  
    }
    \label{fig:cr_population_ne}
\end{figure*}

\Fref{fig:cr_population_ne}~(a) shows examples of $R_1^p(n_e, T_e, T_v)$ with 
$T_\mathrm{e} = 10 \mathrm{\;eV}$ and    
$T_\mathrm{v} = 0.1 \mathrm{\;eV}$, for several different values of \Ne.
Horizontal position shows the excited states of hydrogen molecule; each column separated by dotted lines show an electronic state, while in each column the vibrational quantum number dependence is shown. 
Note that for this calculation, we ignore the quenching and electron-attachment processes. 
MCCC excitation rates are used and Mewe's approximation for the dipole transition where MCCC is unavailable.

Different \Ne-dependence can be seen for different electronic states. 
For example, $R_1^p$ for \Dlevpm states shows little \Ne-dependence, while that for \EFlev state shows a significant \Ne dependence.
Such population dynamics can be understood from the dominant elementary process for each state.

\begin{figure*}
    \includegraphics[width=18cm]{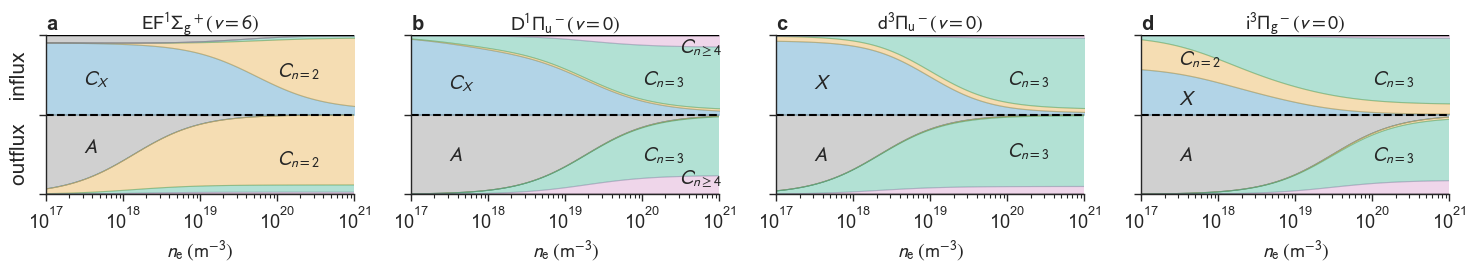}
    \caption{%
        Compositions of the population influx and outflux for some levels, as a function of \Ne. 
        $C_{X}$ indicates the influx from the ground state, while $C_{n=2}$ and $C_{n=3}$ indicates flux from / to $n=2$ and $n=3$ states, respectively.
        In the lower density region, the outflux from all of these states is the radiative decay (indicated by $A$) to lower states, while in higher density region collisional transition dominates the outflux.
        The critical value of \Ne is different for these states depending on the relative importance of the radiative decay and the collisional transition.
        All the values were calculated with our CRM under \Te= 10 eV and $T_v = 0.1$ eV.
    }
    \label{fig:flux}
\end{figure*}

\Fref{fig:flux} shows the elementary-process composition of influx and outflux.
In the low-\Ne limit, the influx to each state is dominated by the collisional excitation from the ground state (denoted by $C_{n=1}$), while the dominant outflux is the radiative decay to the lower state (denoted by $A$ in \fref{fig:flux}). 
Thus, the population can be written as
\begin{align}
    n_p \approx \frac{
        \sum_v C_{p\leftarrow v}n_{\mathrm{X}}(v)
    }{
        \sum_q A_{q\leftarrow p}
    } n_\mathrm{e},
\end{align}
where $R_1^p \propto n_\mathrm{e}^0$.
This phase is called \textit{coronal} phase, and the population coefficient has no \Ne-dependence.

As \Ne increases, the dominant outflux changes to the collisional transition to other states ($C_{n=2}$ and $C_{n=3}$ in \fref{fig:flux}).
In this region, the population may be written as follows;
\begin{align}
    n_p \approx \frac{
        \sum_v C_{p\leftarrow v} n_{\mathrm{X}}(v) n_\mathrm{e}
    }{
        \sum_q C_{q\leftarrow p} n_\mathrm{e}
    },
\end{align}
where the dominant influx (numerator) and outflux (denominator) are both the electron-impact excitation. 
Thus, the \Ne-dependences on the numerator and denominator are canceled out and $R_1^p \propto n_\mathrm{e}^{-1}$. 
This phase is called \textit{saturation} phase.

Since \Dlevpm has shorter radiative lifetime ($\approx 3$ ns), the critical value of \Ne between the \textit{coronal} to \textit{saturation} phases is higher.
At \Ne$\approx 10^{19}\;\mathrm{m^{-3}}$, this state is still in \textit{coronal} phase and thus the population coefficient has no \Ne-dependence (see \fref{fig:cr_population_ne}).
On the other hand, since \EFlev has longer radiative lifetime ($\approx 100$ ns), this transition happens in a lower \Ne, and above this critical density, $R_1^p \propto n_\mathrm{e}^{-1}$.

Lower panels in \fref{fig:cr_population_ne} shows the \Ne-dependence of the population coefficient for several electronic states (vibrational populations are summed up). 
In the low-density region $R_1^p \approx n_\mathrm{e}^0$ while in high-density regions  $R_1^p \approx n_\mathrm{e}^{-1}$, which is consistent with the above discussion.

$R_1^p$ values for \glev and \ilevpm levels show a positive \Ne-dependence in a certain density range in contrast to those for the other levels.
This can be understood from the contribution of the two-step excitation ($C_{n=2}$-influx in \fref{fig:flux}).
Since $(n=2)$-state population ($n_2$) is proportional to $n_\mathrm{e}^1$ in the low-density region, the two-step excitation flux is proportional to
$n_2 n_\mathrm{e} \propto n_\mathrm{e}^2$. 
Because of the contribution of this process, $R_1^p$ values can have a positive \Ne-dependence.
The population converges to $n_p\propto n_\mathrm{e}^0$ at the high-density limit as similar to the other states.

\subsection{Dependence on different datasets}

\subsubsection{Difference due to the rate-approximation methods}

Current database, including MCCC, do not provide the cross-sections \textit{from} the excited states with $n\geq 3$, and there are uncertainty which approximations are more appropriate, i.e., Mewe's approximation (Sec.~\ref{subsubsec:mewe}) or helium approximation (Sec.\ref{subsubsec:helium}).
Two solid curves in each lower panel of \fref{fig:cr_population_ne} show the population coefficients predicted by our CRM, but with (bold) Mewe's approximation and (thin) helium-crosssections for unavailable transitions in MCCC database.
As in the higher-\Ne limit the influx and outflux are dominated by the electron-impact transitions among excited states (see \fref{fig:flux} too), the difference in the values of $R_1^p$ becomes bigger. 
On the other hand, in the lower-\Ne limit the influx is the excitation from the ground state (the cross sections of which are available in the MCCC database~\cite{mccc}) while the dominant outflux is the radiative decay. 
Thus the uncertainty due to the cross-sections for excited states is smaller.
We will examine this effect later.

\subsubsection{The effect of the dissociative attachment}

Thin dotted curves in \fref{fig:cr_population_ne} shows the same calculation but with the dissociative attachment process included.
Since the proposed rate coefficients for the dissociative attachment for $n=3$ states in Ref.~\cite{yacora} is much higher than the electron-impact excitation rate (i.e., $\sum_q C_{q\rightarrow p} \gg \sum_k C_{k\rightarrow p}^\mathrm{attach}$), all the $(n=3)$-states, including \Dlevpm, fall in the \textit{saturation} phase at $n_\mathrm{e} > 10^{17} \mathrm{\;m^{-3}}$ if this process is included.

This effect may be examined by a \Ne-dependence of the population ratio, e.g., that between \EFlev and \Dlevpm.
If the proposed rate is correct and both the states are in the \textit{saturation} phase, the population ratio should have only a weak \Ne-dependence.
On the other hand, if the proposed rate is overestimated, \Dlevpm is in \textit{coronal} phase and thus the population ratio should have a strong \Ne-dependence.
This will be discussed later in Sec.~\ref{subsec:LHD} and \fref{fig:ratio_distribution}.

 
\subsubsection{The difference between the excitation-rate datasets}

\begin{figure*}
    \includegraphics[width=18cm]{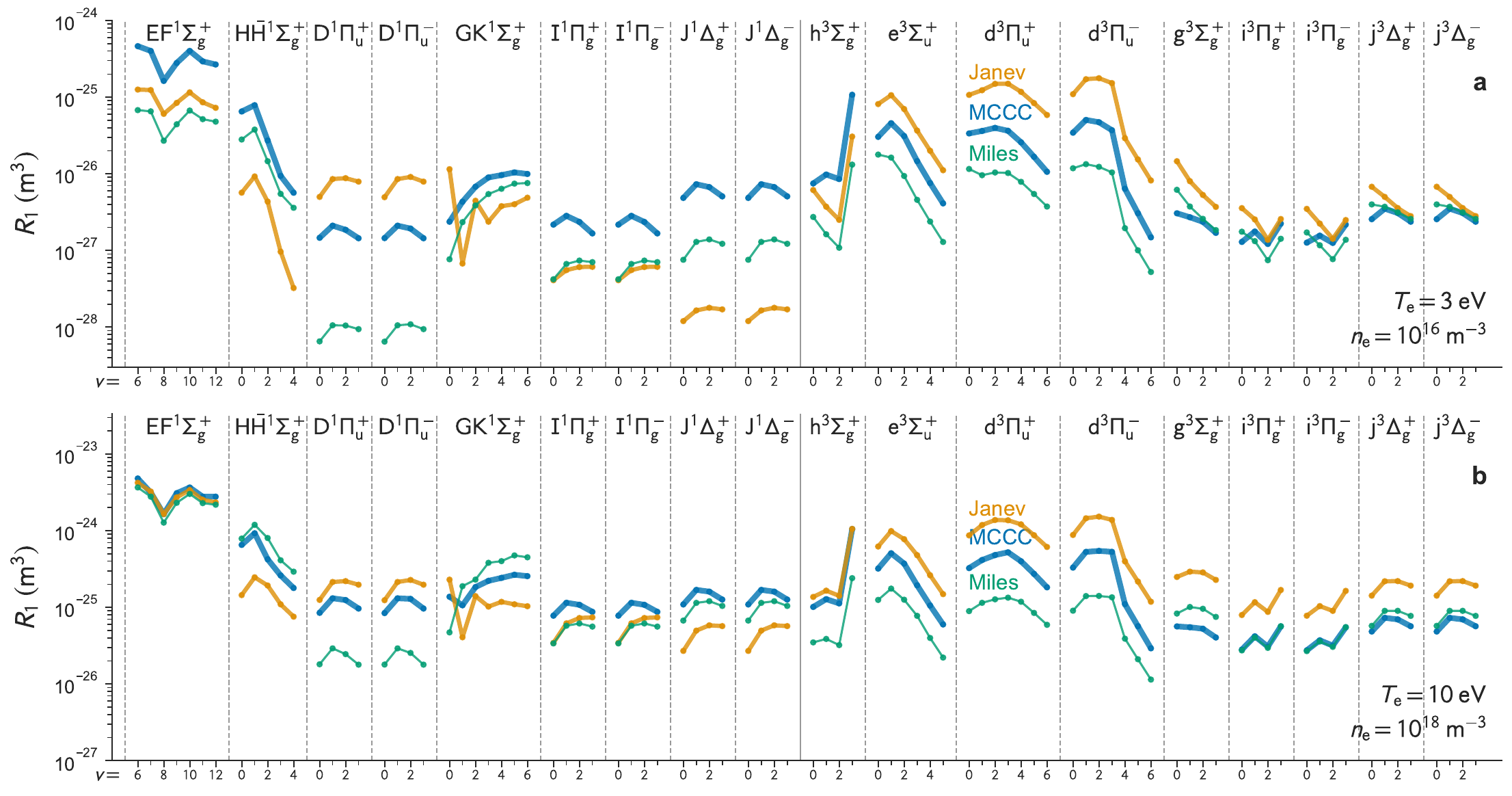}
    \caption{%
        Population coefficients $R_1^p$ for some excited states of molecular hydrogen calculated with our CRM, with different datasets for the electron-impact transitions.
        Top and bottom panels show the prediction under different values of \Ne and \Te.
        By different colors, the results with different datasets (MCCC~\cite{mccc}, Janev~\cite{Janev2003-ca}, and Miles~\cite{Miles1972-md}) are shown.
    }
    \label{fig:cr_population_rates}
\end{figure*}

In \fref{fig:cr_population_rates}, we show the results with different sets of the excitation rates, i.e., those by Janev (orange), Miles (green), and MCCC (blue), without including the dissociative attachment process.
Because of the difference in the rate coefficients for electron-impact transitions, the population coefficients are different.
In higher-temperature plasma (\fref{fig:cr_population_rates}~(a)), the population by Janev and MCCC are similar, while that by Miles shows a significant difference from the other two.
On the other hand, in lower-temperature plasma, the population in \EFlev show a significant difference among these three datasets.

\section{Experimental comparisons\label{sec:experiment}}

In order to compare with the CRM predictions described above, we carry out experimental observations of emission spectra from molecular hydrogen.
To cover a wide range of the parameter space, we measure the emission from a low-density RF plasma in Shinshu university (Sec.~\ref{subsec:shinshu}), and a divertor region of Large Helical Device (LHD, Sec.~\ref{subsec:LHD}).
As described below, the RF plasma has $n_\mathrm{e}\approx 10^{16} \mathrm{\;m^{-3}}$ and $T_\mathrm{e}\approx 10^0\mathrm{\;eV}$, while the LHD divertor has much higher temperature and density range of $n_\mathrm{e}\approx 10^{19} \mathrm{\;m^{-3}}$ and $T_\mathrm{e}\approx 10^1\mathrm{\;eV}$.
Furthermore, the parameter in the LHD divertor can be varied by changing the heating and fueling conditions. 
This even widens the parameter space we consider here.

\subsection{Low-density plasma experiment in Shinshu university\label{subsec:shinshu}}

We measured an emission spectrum from a low-temperature RF plasma at Shunshu University, Japan. 
A hydrogen plasma was generated in a vacuum chamber made of Pyrex glass, with 50-mm inner diameter and 1100-mm length. 
This chamber is located inside two solenoids, which generates 0.012 T magnetic field at the plasma center. 
A $\approx$100-W RF power is applied to generate the plasma.
This experimental setup is similar to that shown in Ref.~\cite{Sawada2010-vl}, 
however instead of the pure-helium plasma in the previous work, we generated a plasma with helium-hydrogen gas mixture  in this work.
The partial gas pressure of hydrogen molecule and helium atom are 0.07 and 0.02 torr, respectively.

An echelle spectrometer with a crossed disperser configuration (EMP-200 Bunko-Keiki, 202-mm focal length) and a charge-coupled-device (CCD) detector (DV420A-OE, Andor inc., $255\times 1024$ pixels, $26\times 26 \mathrm{\;\mu m^2}$ pixel size) were used to measure the emission spectrum.
This spectroscopic system can simultaneously measure the spectrum in the wavelength range of 400 -- 800 nm with 0.05-nm resolution.

The values of \Ne and \Te in the plasma were estimated as $T_\mathrm{e} \approx 3\;\mathrm{eV}$ and $n_\mathrm{e} \approx 10^{16}\;\mathrm{m^{-3}}$,
based on the helium-line-ratio method described in Ref.~\cite{Sawada2010-vl}.
In this method, the absolute emission intensities from multiple excited states of helium atoms (
    $3s\,^1\mathrm{S}$, $3p\,^1\mathrm{P}$, $3d\,^1\mathrm{D}$,
    $4s\,^1\mathrm{S}$, $4p\,^1\mathrm{P}$, $4d\,^1\mathrm{D}$,
    $5s\,^1\mathrm{S}$, $5d\,^1\mathrm{D}$, $6d\,^1\mathrm{D}$, 
    $3s\,^3\mathrm{S}$, $3p\,^3\mathrm{P}$, $3d\,^3\mathrm{D}$,
    $4s\,^3\mathrm{S}$, $4d\,^3\mathrm{D}$,
    $5s\,^3\mathrm{S}$, and $5d\,^3\mathrm{D}$) 
are fitted by a prediction of collisional-radiative model for helium atoms. 
By the fit, several adjustable parameters are determined, which are 
\Ne, \Te, density of metastable states ($2s ^1\mathrm{S}$ and $2s ^3\mathrm{S}$), and the radiative reabsorption effect by $3p ^1\mathrm{P}$ and $4p ^1\mathrm{P}$ states. 
Note that although many adjustable parameters are used to fit multiple emission intensities, the analysis is robust enough.
The value of \Te is basically determined by the population ratio among 
$3s\,^1\mathrm{S}$, $4s\,^1\mathrm{S}$,
$3s\,^3\mathrm{S}$ and $3s\,^3\mathrm{P}$ states, where the other effects (excitation from metastable and radiation reabsorption) are negligible in this condition.  
The value of \Ne is in turn determined by comparing the absolute intensities with the helium atom density and the excitation rates, which are the function of \Te. 

\begin{figure*}
    \includegraphics[width=18cm]{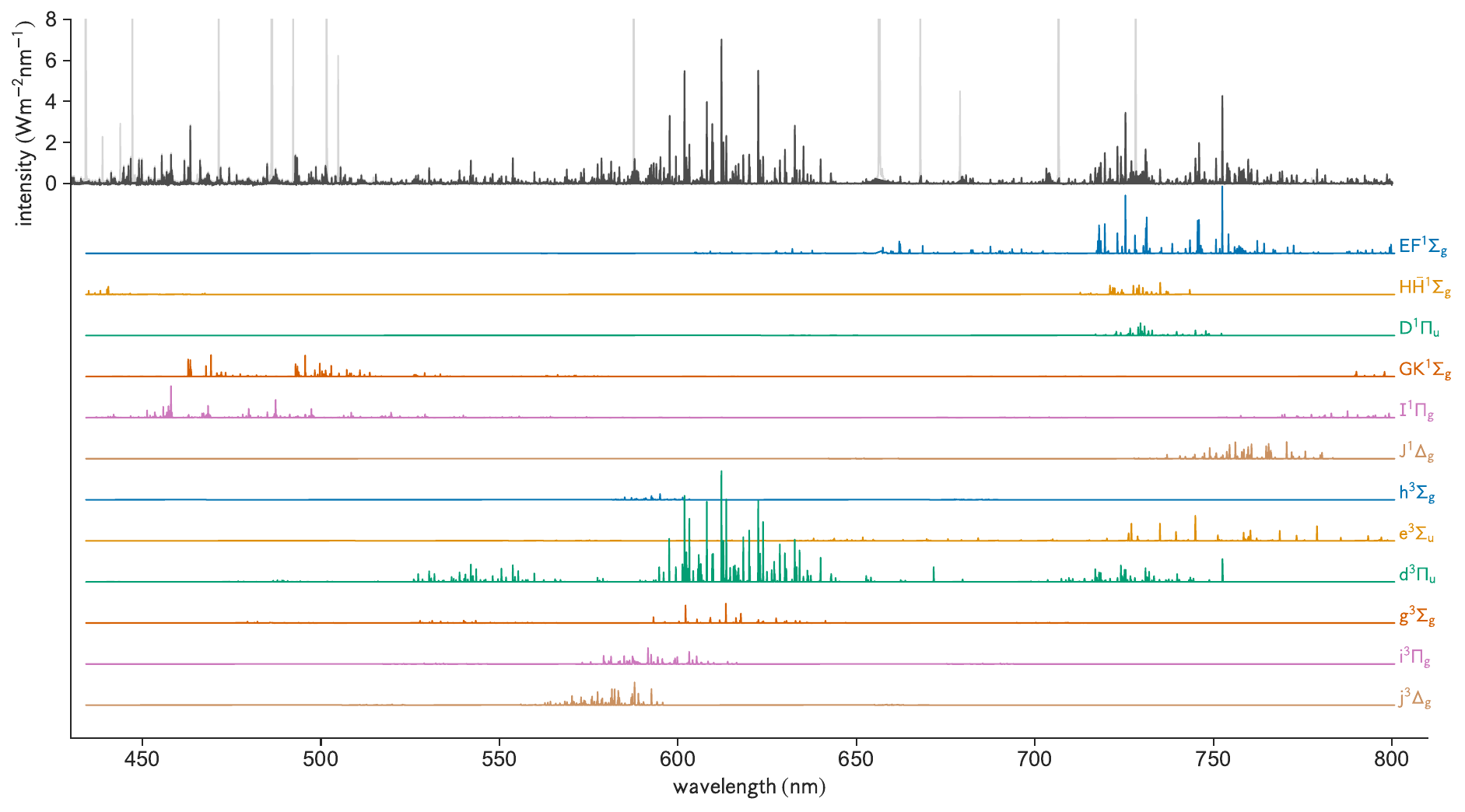}
    \caption{%
        (top) Emission spectrum observed from the low-density RF plasma with helium-hydrogen mixture. 
        To clearly present the emission lines from molecular hydrogen, strong atomic lines and helium lines are omitted.
        (bottom) The composition of the hydrogen-molecule spectrum for each upper-electronic states, which is obtained from the spectral fit. 
    }
    \label{fig:spectrum_sawada}
\end{figure*}

\Fref{fig:spectrum_sawada} shows the observed emission spectrum. 
To clearly present the molecular emission, strong atomic helium and hydrogen emission lines as well as continuum emission were excluded from the figure.
Three clusters of lines can be found in the visible range; 450--500 nm, 570--630 nm, 720--800 nm.

\subsubsection{Analysis method}

The observed emission corresponds to transitions with the change in electronic, vibrational and rotational states, while the CRM we have developed in Sec.~\ref{sec:crm} resolves only the electronic and vibrational states.
In order to obtain the population from the experiment in each electronic and vibrational state but integrated over the rotational states, we assume the Boltzmann distribution for the rotational population,
\begin{align}
    n_{\alpha'v'N'} \approx \frac{n_{\alpha'v'}}{Z_{\alpha'v'}}
    (2S' + 1)(2 N' + 1) g^\mathrm{as}_{N'}
    \exp\left(-\frac{E_{\alpha'v'}(N')}{kT_{\alpha'v'}}\right),
\end{align} 
where $\alpha'$, $v'$, $S'$, and $N'$ indicate the electronic state, vibrational quantum number, electron spin quantum number, and rotational quantum number of the upper level, respectively, and $E_{\alpha'v'}(N')$ is the excitation energy.
$g^\mathrm{as}_{N'}$ is the nuvlear spin statistical weight.
$Z_{\alpha',v'}$ is the partition function.
We assume an independent rotational temperature $T_{\alpha',v',N'}$ for each electronic and vibrational state.
The population and rotational temperature in each electronic and vibronic state, $n_{\alpha',v'}$, were optimized so that the predicted emission intensity $I^{\alpha'v'N'}_{\alpha''v''N''} = A^{\alpha'v'N'}_{\alpha''v''N''} n_{\alpha'v'N'}$ represents the observed spectrum the best.
We use the literature values for the A coefficient if available, otherwise we use the H\"onl-Londom approximation assuming Hund's case b with the vibrationally-resolved A coefficient values~\cite{Fantz2006-cn}.

We fit the entire spectrum based on the above assumption. 
The decomposition of the spectrum is shown in the bottom part of \fref{fig:spectrum_sawada}. 
It is clear that the three clusters have different upper-electronic states; 
450 -- 500 nm lines are mainly originated from \GKlev and \Ilevpm states, 
570 -- 630 nm lines are from \dlevpm, \ilevpm, and \jlevpm states, and 
720 -- 800 nm lines are from \EFlev, \Hlev, \Dlevpm, \Ilevpm, and \Jlevpm states.

\begin{figure*}
    \includegraphics[width=18cm]{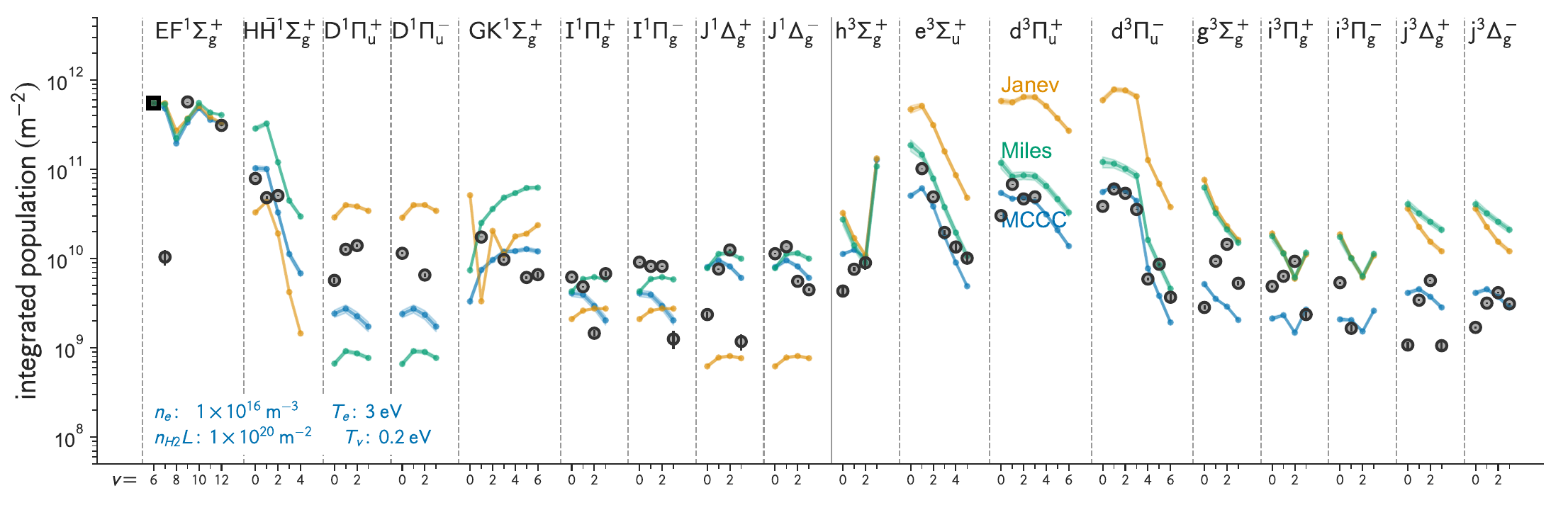}
    \caption{%
        Distribution of the line-integrated population of molecular hydrogen observed for the low-density RF plasma~(markers). 
        Kinked lines for each electronic state are the prediction by the CRM with three different excitation cross section (blue: MCCC, green: Miles, orange: Janev).
        The values of \Ne and \Te to calculate the population were estimated based on the helium line-ratio method~\cite{Sawada2010-vl}, which are 
        $T_\mathrm{e}\approx 3\;\mathrm{eV}$ and $n_\mathrm{e}\approx 10^{16}\;\mathrm{m^{-3}}$.
        \Eref{eq:Tv} is used to compute $T_v \approx 2400$ K.
        The only one adjustable parameter is the line-integrated molecular density, $n_\mathrm{H_2}L$, is adjusted so that the population in \EFlev$(v=6)$ is identical between the observation and the prediction, which are $n_\mathrm{H_2}L \approx 10^{20}\;\mathrm{m^{-2}}$.
    }
    \label{fig:population_sawada}
\end{figure*}

Markers in \fref{fig:population_sawada} shows the population obtained from the analysis.
We also plot the prediction by our CRM with the values of \Ne and \Te obtained from the helium-line-ratio method.
The value of $T_v$, which is the vibrational temperature of the ground state molecules, are obtained from the empirical relation~\cite{Brezinsek2002-ss},
\begin{align}
    \label{eq:Tv}
    \frac{T_v}{\mathrm{K}} \approx 2400 + 2.6\times 10^{-16} \left[\frac{n_\mathrm{e}}{\mathrm{m^{-3}}}\right].
\end{align}
For this low-density plasmas, this means essentially $T_v\approx 2400\mathrm{\;K}$.
Only one adjustable parameter in the comparison is the value of $n_\mathrm{H_2}L$ (where $L$ is the effective diameter of the plasma). 
We scale the prediction by adjusting this value so that the population in \EFlev$v=6$ states becomes identical between the experiment and the simulation (the square marker in \fref{fig:population_sawada}).
From this scaling, $n_\mathrm{H_2}L \approx 10^{20}\;\mathrm{m^{-2}}$ is obtained, which is consistent with the gas density ($\approx 10^{22}\;\mathrm{m^{-3}}$) and the plasma column diameter $L\approx10^{-2}\;\mathrm{m}$.

The three types of lines in \fref{fig:population_sawada} show the CRM predictions with different cross-section datasets, i.e., those by MCCC (blue), by Miles (green), and Janev (orange).
It is clearly seen that the prediction by the MCCC cross-sections represents the experimental observation the best, in particular the populations in \EFlev, \Hlev, \Jlevpm, \hlev, \elev, \dlevpm, and \jlevpm.
Janev and Miles cross-sections particularly overestimate the triplet population. 

However, even the prediction by MCCC shows a significant underestimation in the population in \Dlevpm states. 
The overpopulation might be caused by the radiation reabsorption, since \Dlevpm states are optically allowed levels from \Xlev states.
From the ground state density ($\approx 10^{22} \mathrm{\;m^{-3}}$), the mean free path of the photon from \Dlevpm states is in the order of $\lesssim 10^{-3}$ m.
This is much shorter than the plasma size of this experiment ($\approx 10^{-2}$ m), and thus this radiation reabsorption effectively decreases the radiative decay rate from \Dlevpm state and increases its population.
Furthermore, the $v$-dependence in \Dlevpm is also consistent with the trend of the Franck-Condon factor with \Xlev, i.e., $q_{\mathrm{X^1\Sigma_g^+}(v=0)-\mathrm{D^1\Pi_u^-}(v=0,1,2)} \approx [0.10, 0.17, 0.18]$; since the emission from $v=2$ state of \Dlevpm has bigger absorption rate than that from $v=0$ state, the overpopulation in $v=2$ state should be more significant.
These are all consistent with the experiment, although this effect is not included in our CRM analysis.

\subsection{LHD experiment\label{subsec:LHD}}
In order to examine the dataset in other parameter spaces, we carried out the emission observation from LHD divertor region, with the same setting described in Ref.~\cite{ishihara}.
The light from the divertor region is collected by an optical lens, introduced into an optical fiber, and guided to the entrance slit of another echelle spectrometer (home-made, 300-mm focal length~\cite{Tanaka2018}). 
This spectrometer can simultaneously measure an emission spectrum in the entire visible range (430 -- 800 nm) with the high-wavelength resolution (0.06 nm).

\begin{figure}
    \includegraphics[width=8.5cm]{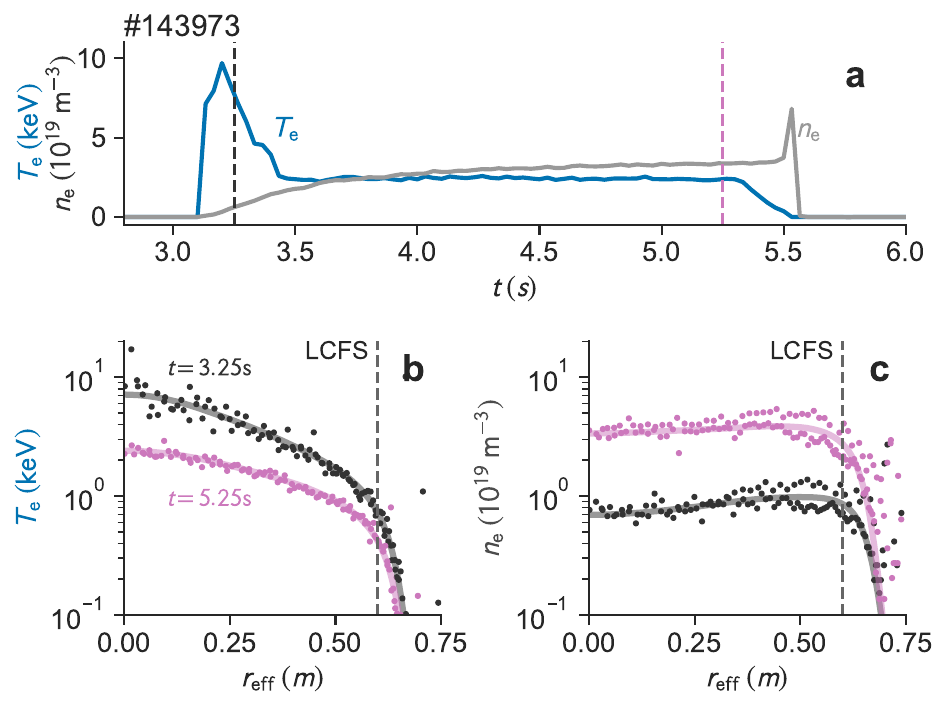}
    \caption{%
        (a) A temporal evolution of \Te and \Ne at the plasma center of the LHD.
        The radial distributions of (b) \Te and (c) \Ne for two different timings $t=3.25$ s and $t = 5.25$ s.
        The position of the LCFS are indicated by the dotted line.
    }
    \label{fig:neTe}
\end{figure}

\Fref{fig:neTe} shows the summary of a typical LHD experiment.
The temporal evolutions of \Te and \Ne on the plasma axis of the LHD are shown in \fref{fig:neTe}~(a), while \fref{fig:neTe}~(b) and (c) show the radial distributions of \Te and \Ne, respectively, for two measurement timings. 
The values of \Te and \Ne are high in the confined region (inside the last closed flux surface, LCFS), while they become much smaller in the divertor.
In Ref.~\cite{Kobayashi2010-iv}, it has been found that the values of \Te and \Ne at the divertor, $T_\mathrm{e}^\mathrm{div}$ and $n_\mathrm{e}^\mathrm{div}$, respectively, approximately have the following relationship with the values on the LCFS,
\begin{align}
    \label{eq:TeNe}
    \frac{T_\mathrm{e}^\mathrm{div}}{\mathrm{eV}} 
    &\approx 30 \times \left[\frac{n_\mathrm{e}^\mathrm{LCFS}}{10^{19}\mathrm{\;m^{-3}}}\right]^{-0.5}\\
    \frac{n_\mathrm{e}^\mathrm{div}}{10^{19}\mathrm{\;m^{-3}}} 
    &\approx 0.08 \times \left[\frac{n_\mathrm{e}^\mathrm{LCFS}}{10^{19}\mathrm{\;m^{-3}}}\right]^{1.5}
\label{eq:nete_empirical}
\end{align}
It is important to note that, besides the above empirical relation, the exact \Ne and \Te values at the emission locations are not available.
Thus, in the following we use this empirical relation with the uncertainty of 30\%.

\begin{figure*}
    \includegraphics[width=18cm]{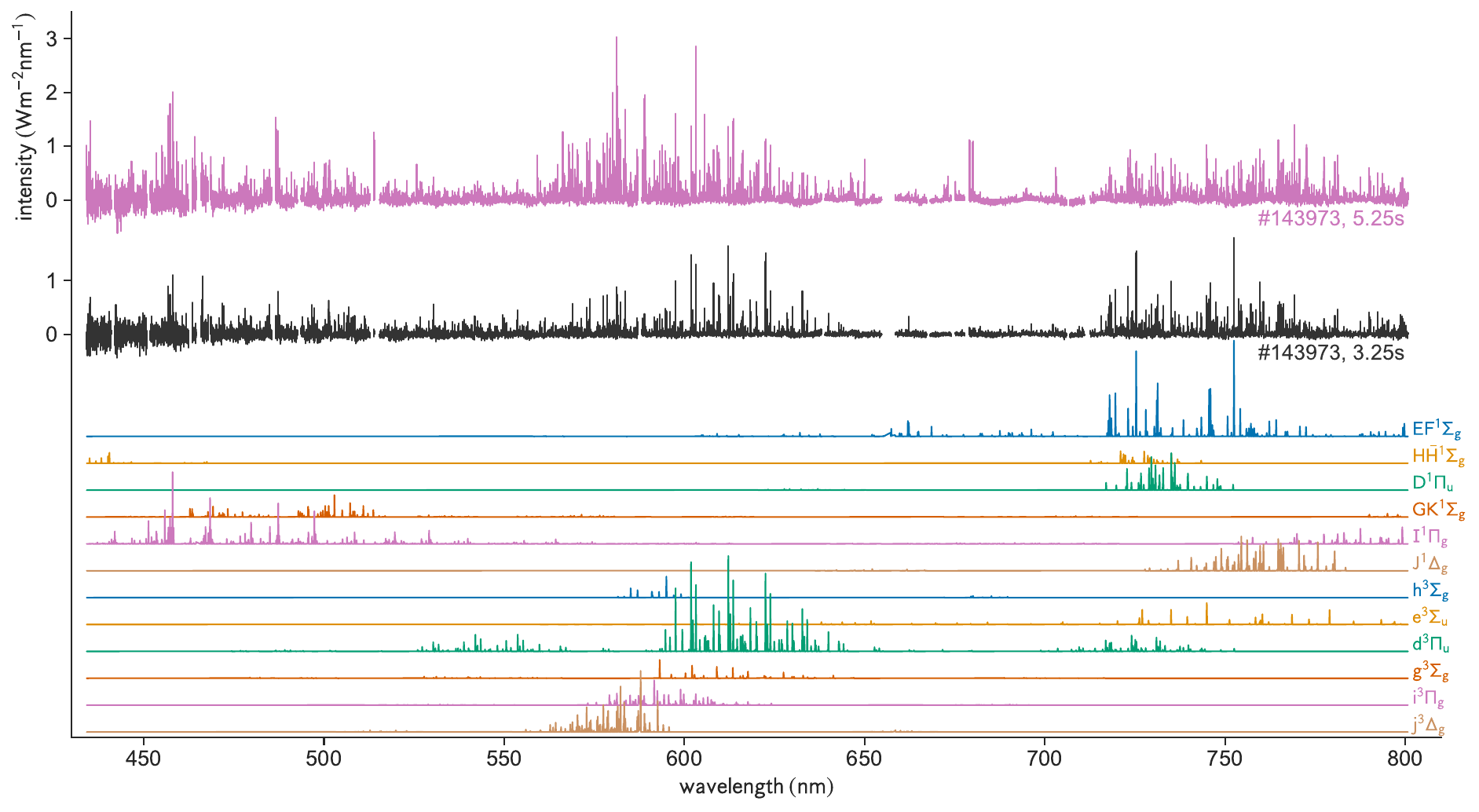}
    \caption{%
        A similar figure to \fref{fig:spectrum_sawada}, but for the LHD divertor plasma.
        (top) Emission spectrum observed at (gray) $t=3.25$ s and (magenta) $t=5.25$ s. 
        To clearly present the emission lines from molecular hydrogen, strong atomic lines are omitted.
        (bottom) The composition of the hydrogen-molecule spectrum observed at $t=3.25$ s for each upper-electronic states, which is obtained from the spectral fit. 
    }
    \label{fig:spectrum}
\end{figure*}

\Fref{fig:spectrum} shows the two emission spectra observed for LHD experiment \#143973 at $t=3.25$~s and $t=5.25$~s.
The shapes of the spectrum are different between the two timings; the two strong lines in 720--800 nm in the $t=3.25$ s spectrum disappear in $t=5.25$~s, while new lines appear in 570--600 nm.

\begin{figure*}
    \includegraphics[width=18cm]{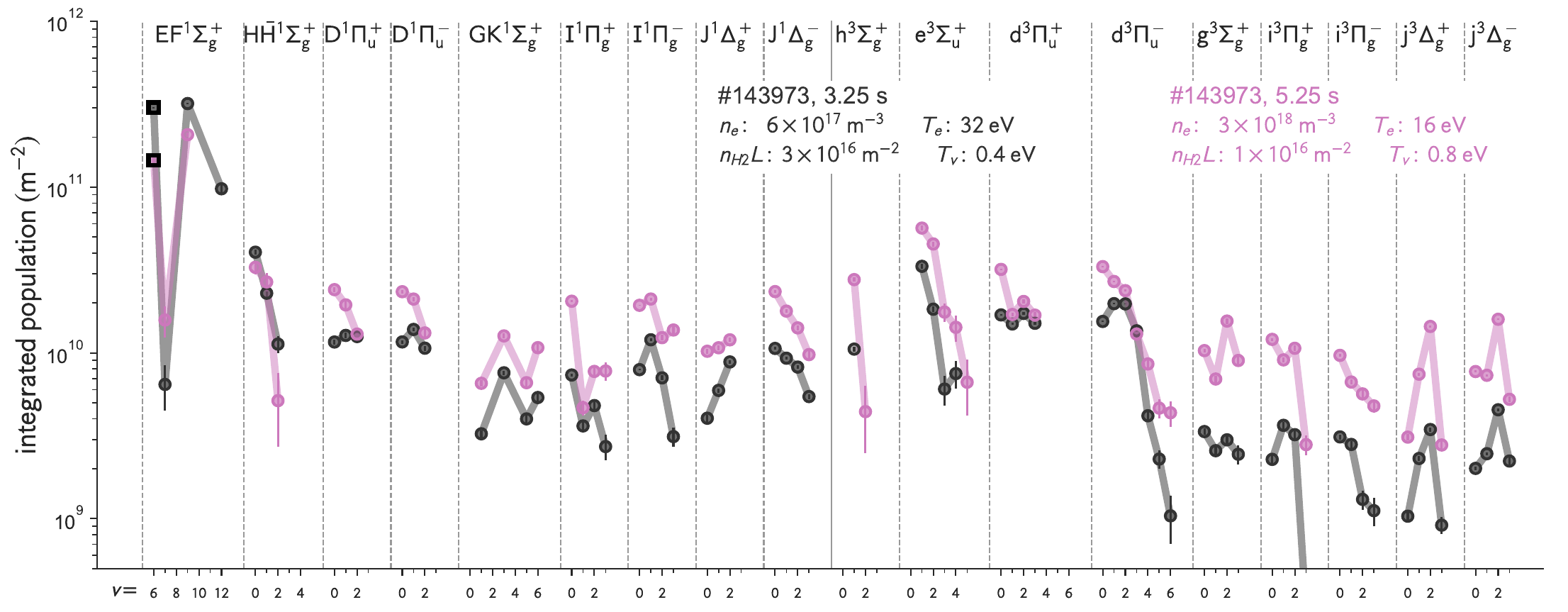}
    \caption{%
        Population distributions observed for the LHD divertor at (grey) $t=3.25$ s and (magenta) $t=5.25$ s.
        Different behavior in each state can be seen, e.g., the population in \EFlev state decreases while those in the triplet states increase.
    }
    \label{fig:population_comparison}
\end{figure*}

The same analysis to that described in the previous section is performed for all the exposure frames in this experiment. 
\Fref{fig:population_comparison} shows the population obtained from the analysis for the two timings. 
From $t=3.25$~s to 5.25~s, the population in \EFlev decreases while those in \glev, \ilevpm, and \jlevpm increase. 
Note that although the values of \Te and \Ne changes during LHD experiments, as the timescale of the parameter change in the plasma is much longer ($\gtrsim 10^{-3}$ s) than that of the population change ($\lesssim 10^{-7}$ s), the quasi steady-state approximation for the excited state population is still valid.
The populations in \Dlevpm and \dlevpm stay almost the same in these two timings.
This is consistent with the parameter dependence on $R_1$ as shown in \fref{fig:cr_population_ne}.

\begin{figure*}
    \includegraphics[width=18cm]{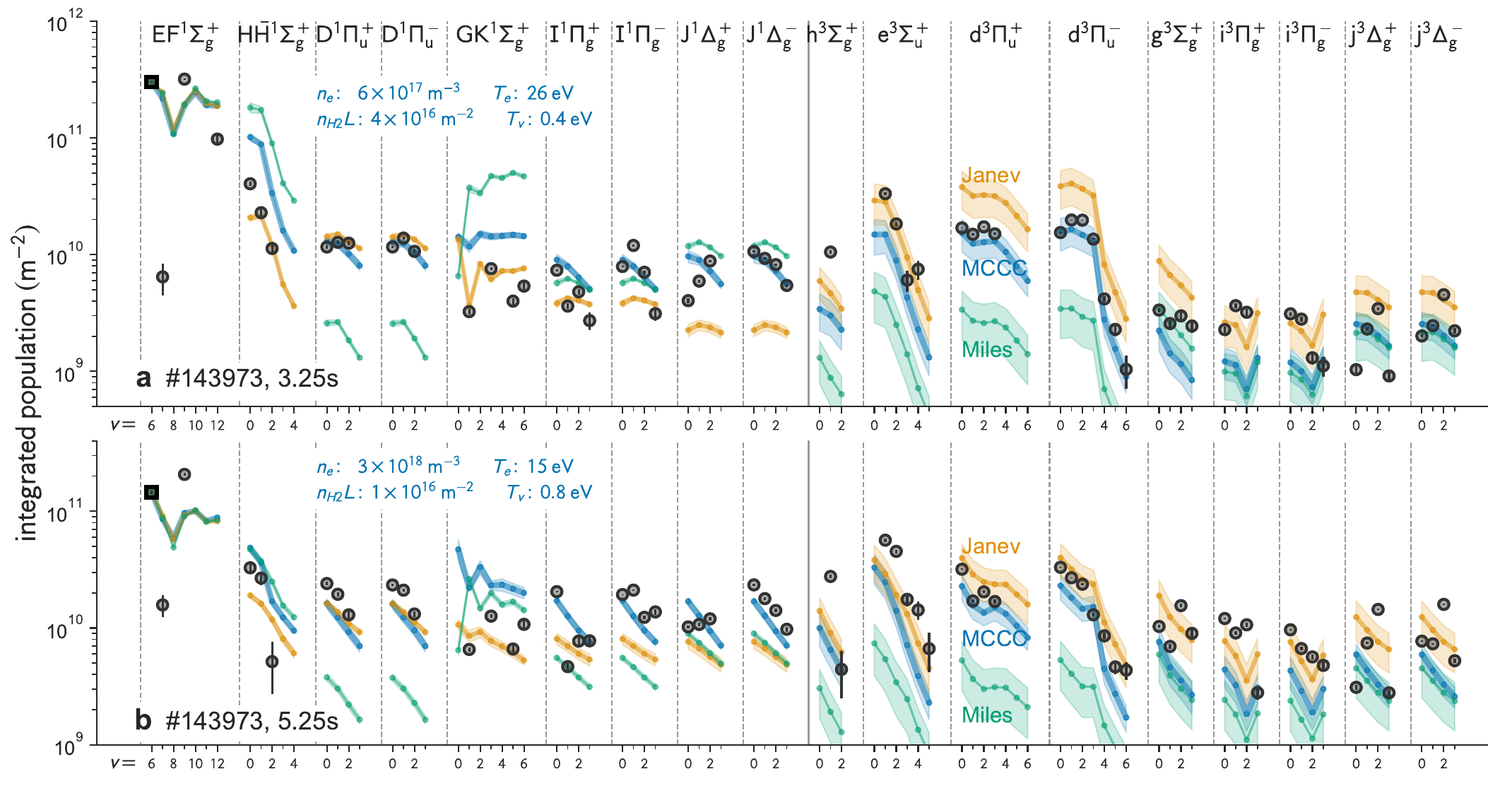}
    \caption{%
    A similar figure to \fref{fig:population_sawada} but for the LHD divertor plasmas.
    Experimentally observed population distribution is shown by markers, while kinked lines are the prediction by the CRM with three different excitation cross section (blue: MCCC, green: Miles, orange: Janev).
    The values of \Ne, \Te, and $T_v$ to calculate the population were estimated based on \eref{fig:neTe} and \eref{eq:Tv}.
    The only one adjustable parameter is the line-integrated molecular density, $n_\mathrm{H_2}L$, is adjusted so that the population in \EFlev$(v=6)$ is identical between the observation and the prediction.
    }
    \label{fig:population}
\end{figure*}

\Fref{fig:population} shows the comparison with our CRM predictions. 
As similar to the discussion for \fref{fig:population_sawada}, we used fixed \Ne, \Te, and $T_v$ values based on \eref{eq:TeNe} and \eref{eq:Tv}.
To account for the uncertainty in these parameters, we assume independent 10\% error bands for these three parameters.
The value of $n_\mathrm{H_2} L$ is estimated so that the \EFlev($v=6$) population by the experiment and predictions match.

The predictions with the cross sections by MCCC and Janev equally well represent the experimental observations. 
Miles' prediction show the significant underestimation on the \Jlevpm, \hlev, and \dlevpm populations.

Even with the CRM using the MCCC dataset, the fit is worse than that in the low-density RF plasma \fref{fig:population_sawada}.
The populations in \GKlev, \ilevpm, and \jlevpm are not well reproduced by the CRM, in particular in the higher density plasma (the lower panel of \fref{fig:population}).  
The reason might be the uncertain cross sections \textit{from} the excited state. 
We will discuss this issue in the next section.

\begin{figure}
    \includegraphics[width=7.5cm]{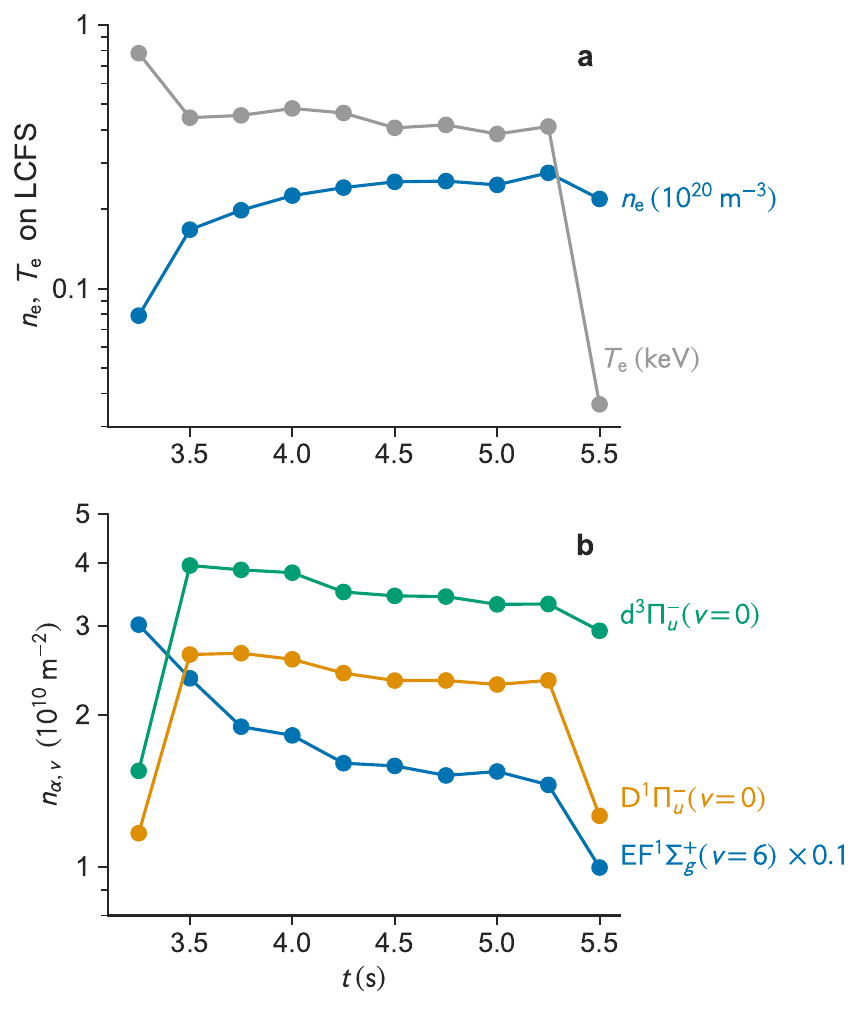}
    \caption{%
        (a) A temporal evolution of \Te and \Ne on the LCFS for LHD experiment \#143973.
        (b) The temporal evpolutions of the populations in 
        \EFlev $(v=6)$, \Dlevm $(v=0)$, and \dlevm $(v=0)$.
        Not only the population but also the population ratio changes in time, according to the change in \Ne and \Te.
    }
    \label{fig:temporal_evolution}
\end{figure}

We also conduct the above emission measurement and analysis for the spectra observed in other timings. 
\Fref{fig:temporal_evolution} shows the temporal evolution of the population of several excited states, for this particular LHD experiment.
The population and the population ratios change in time.
In the first frame where the plasma has the lowest density and highest temperature during the discharge, the population in \EFlev is highest while that in \Dlevm is lowest.
While \Ne increases as time goes by, the \EFlev population decreases while \Dlevm population increases.
At the final frame, where the temperature dropped significantly, the \Dlevm population decreases significantly while the \dlevm population stays almost the same.
This is consistent with the energy dependence of the cross sections; the cross section to the triplet state has higher efficiency at the lower temperature (see \fref{fig:crosssections} and \fref{fig:rate_comparison}).

\begin{figure*}
    \includegraphics[width=18cm]{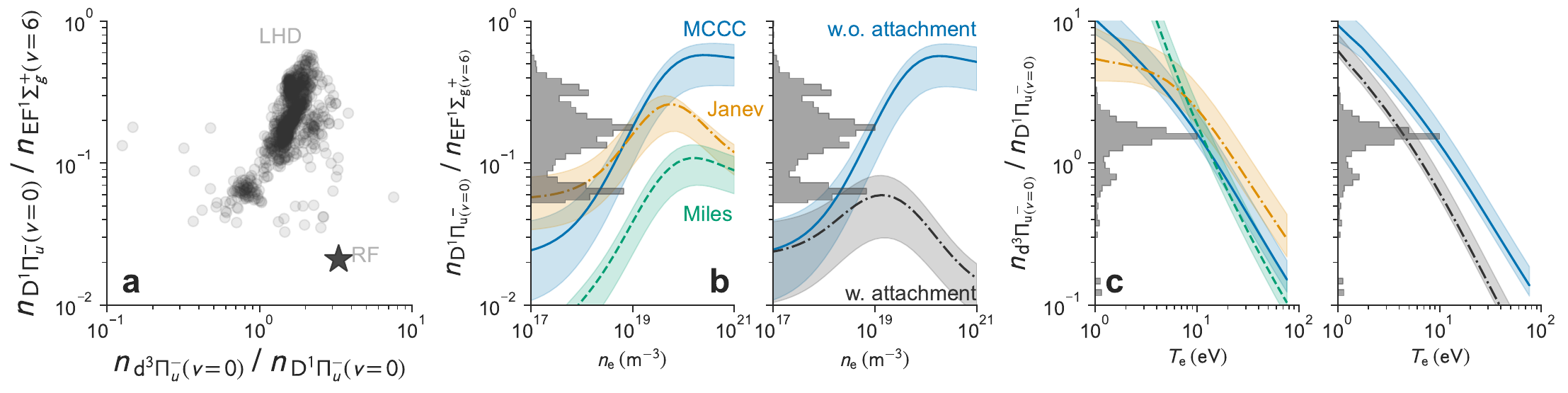}
    \caption{%
    A distribution of population ratios observed for the LHD divertor and the low-density RF plasma. 
    (a) A correlation between the two population ratios, \EFlev$(v=6)$~/~\Dlevm$(v=0)$ and \dlevm$(v=0)$~/~\Dlevm$(v=0)$.
    (b, c) Histograms of the observed population ratios and the predicted population ratios by our CRM.
    On the left of each panel, the predictions by the three different cross-section datasets (blue: MCCC, green: Miles, orange: Janev) are shown.
    On the right, we compare the predictions by the CRM (blue) without and (black) with the recommended electron-attachment rates.    
    We assume $T_v = 0.2$--$1.0\;\mathrm{eV}$ and (b) $T_\mathrm{e} = 5$--$20\;\mathrm{eV}$ and (c) $n_\mathrm{e} = (0.5$--$2)\times 10^{19}\;\mathrm{m^{-3}}$ for the CRM.
    }
    \label{fig:ratio_distribution}
\end{figure*}

The same analysis are repeated for multiple LHD experiments spanning wide range in the parameter space.
\Fref{fig:ratio_distribution} shows the distribution of some population ratios. 
The star markers in the figure shows the same population ratios from the low-temperature RF plasma. 

In \fref{fig:ratio_distribution}~(b) and (c) we compare the observed population ratios and the CRM prediction.
On the left side of each panel, the CRM prediction with the three different excitation cross-sections datasets are shown in different colors.
On the right half of the panels, the CRM predictions with and without the dissociative electron-attachment are plotted.
Note that we assume $T_v = 0.2$--$1.0\;\mathrm{eV}$ and (b) $T_\mathrm{e} = 5$--$20\;\mathrm{eV}$ and (c) $n_\mathrm{e} = (0.5$--$2)\times 10^{19}\;\mathrm{m^{-3}}$ for the CRM.

The population ratios between \EFlev and \Dlevm shows a sensitivity on \Ne, as expected from the discussion in Sec.\ref{sec:crm_dynamics}.
The CRM with the Miles' dataset or Janev's dataset fail to reproduce the range of the observed population ratios.
This \Ne-sensitivity disappears if we include the electron-attachment process, and fails to reproduce the observed population ratios.

The population ratios between \dlevm and \Dlevm shows a \Te-sensitivity.
Although this is not perfectly independent from \Ne, this suggests that this ratio might be useful to diagnose the value of \Te from the molecular emission spectrum.

\section{Summary and conclusions}

We developed a CRM for hydrogen molecules, and discussed the effects of different datasets.
The CRM prediction was compared with two different experiments, which is the lower-density RF plasmas and higher-density LHD divertor plasmas.
From the population comparisons the following is found;
\begin{enumerate}
    \item Compared with Miles' and Janev's cross sections, MCCC cross sections show a better agreement with the experiment.
    \item CRM prediction is more accurate in the lower density plasma, where the excitation \textit{from} excited states is negligible. In the higher density plasmas this effect is more significant. Cross-sections from excited states might be necessary to improve for the high density plasma diagnostics. 
    \item Dissociative electron-attachment rate proposed in Ref.~\cite{yacora} might be too high. 
    \item Some pairs of the emission lines of hydrogen molecule show \Ne and \Te dependencies, e.g., \EFlev-\Dlevpm has a \Ne dependence, and \Dlevpm-\dlevpm has a \Te dependence. This may be useful for the plasma diagnostics for low-temperature hydrogen plasmas.
\end{enumerate}

Since most of the important data is already available for other isotopologues, such as $\mathrm{D}_2$, the same analysis can be done also.
This is left for the future study.

\begin{acknowledgments}
    This work was partly supported by the U.S. D.O.E contract DE-AC05-00OR22725, the Australian Government through the Australian Research Council's Discovery Projects funding scheme (project DP240101184), and by the United States Air Force Office of Scientific Research. 
    K.F. would like to specifically acknowledge Oak Ridge National Laboratory's Laboratory Directed Research and Development program Project No. 11367.
    L.H.S is the recipient of an Australian Research Council Discovery Early Career Researcher Award (project number DE240100176) funded by the Australian Government. 
    HPC resources were provided by the Pawsey Supercomputing Research Centre and the National Computational Infrastructure, with funding from the Australian Government and the Government of Western Australia, and the Texas Advanced Computing Center (TACC) at the University of Texas at Austin. 
    M.C.Z would like to specifically acknowledge the support of the Los Alamos National Laboratory (LANL) ASC PEM Atomic Physics Project. 
    LANL is operated by Triad National Security, LLC, for the National Nuclear Security Administration of the U.S. Department of Energy under Contract No. 89233218NCA000001.
    M.C.Z. would like to specifically acknowledge Los Alamos National Laboratory's Laboratory Directed Research and Development program Project No. 20240391ER.
\end{acknowledgments}

\renewcommand{\thefigure}{A\arabic{figure}}
\setcounter{figure}{0}
\renewcommand{\thetable}{A\arabic{table}}
\setcounter{table}{0}
\renewcommand{\theequation}{A\arabic{equation}}
\setcounter{equation}{0}


\bibliography{refs}

    
\end{document}